\documentclass[10pt, aps, showpacs, preprintnumbers, prd, nofootinbib, preprint, twocolumn]{revtex4}
\usepackage{feynmf}
\usepackage{amsmath}
\usepackage{graphicx}
\usepackage{mathrsfs}
\usepackage{stmaryrd}
\usepackage{amsfonts}
\usepackage{wrapfig}
\usepackage{subfig}
\usepackage{float}
\usepackage{bbm}

\begin{document}
\def\sech{\mathop{\rm{sech}}\nolimits}
\def\arcsinh{\mathop{\rm{arcsinh}}\nolimits}

\title{Large lepton mixing angles from a 4+1-dimensional $SU(5)\times{}A_{4}$ domain-wall braneworld model}
\author{Benjamin D. Callen}\email{bdcallen@student.unimelb.edu.au}
\affiliation{ARC Centre of Excellence for Particle Physics at the Terascale, School of Physics, The University of Melbourne, Victoria 3010, Australia}
\author{Raymond R. Volkas}\email{raymondv@unimelb.edu.au}
\affiliation{ARC Centre of Excellence for Particle Physics at the Terascale, School of Physics, The University of Melbourne, Victoria 3010, Australia}

\begin{abstract}

 We propose an extension of the 4+1D $SU(5)$ domain-wall braneworld of Davies, George and Volkas which includes the addition of a discrete $A_4$ flavor symmetry. We show that lepton mixing and light Majorana neutrino masses can be generated from the additional $A_{4}$ physics while at the same time sufficient parameter freedom can be maintained in the charged fermion sector to produce charged fermion masses and quark mixing naturally from the split fermion mechanism. Importantly, we show that the vacuum realignment problem typical of discrete flavor symmetry models of quark and lepton mixing can be solved by separating the appropriate flavons in the extra dimension, leading to exponentially sensitive suppression of the operators responsible for vacuum realignment.

\end{abstract}

\maketitle

\newpage

\section{Introduction}
\label{sec:introduction}

  The fermion mass hierarchy problem and the origins of quark and lepton mixing are three of the most outstanding problems of the Standard Model. The first concerns the 12 orders of magnitude spread amongst the Standard Model (SM) fermions, from the mass of the top quark down to the (upper bounds of the) masses of the neutrinos. The last two concern the Euler angles and CP phases of the unitary matrices which describe the relative strengths of the charged current interactions amongst the different mass eigenstates, which are the CKM matrix in the case of quarks and the PMNS matrix for leptons. For the CKM matrix, the Euler angles are approximately $\theta_{12}=13^{\circ}$, $\theta_{13}=0.2^{\circ}$, and $\theta_{23}=2.4^{\circ}$ \cite{pdg2010}, while the CP phase is about $\delta_{13}=1.2$. It is known that the PMNS matrix is approximately tribimaximal, with $\theta_{12}=34.06^{+1.16\circ}_{-0.84}$ and $\theta_{23}=45^{\circ}\pm{}7.1^{\circ}$ \cite{pdg2010}. Furthermore, recent experiments, in particular the results from Daya Bay and RENO \cite{dayabaytheta13, renotheta13}, support a small $\theta_{13}$ for the PMNS mixing matrix while the nature of CP violation in the lepton sector is unknown. 

  Extra-dimensional approaches to flavor are a commonly used for the fermion mass hierarchy problem in particular. In extra-dimensional approaches to flavor, the hierarchy of Yukawa couplings to the Higgs is normally generated from overlaps of fermion profiles which are exponentially dependant on the input parameters, an example of which is the split-fermion mechanism of Arkani-Hamed and Schmalz \cite{splitfermions}. 

  Given that there is no known reason for the apparent $3+1$-dimensional nature of our universe, it is interesting to consider whether there could be extra spatial dimensions which are hidden with respect to the known four. In some extra-dimensional theories the hidden dimensions are compactified and miniscule. On the other hand, in braneworld scenarios, our universe is confined to a $3+1$-dimensional \emph{brane} in a higher dimensional space. Braneworld models first became popular with the advent of models with large extra dimensions which could solve the hierarchy problem, most notably the ADD model of Arkani-Hamed, Dimopolous and Dvali \cite{originalbranepaper}. Later, Randall and Sundrum showed that the hierarchy problem could be solved in the RS1 model, in which there are two branes embedded in a slice of Anti-de Sitter space, with a UV brane on which gravity is localized and is strong and an IR brane on which the Standard Model (SM) fields are localized and gravity is weak \cite{randallsundrum1}. The same two authors also proposed an alternative model, the RS2 model, in which the extra dimension of the warped spacetime was in fact infinite with gravity and the SM fields localized on a single fundamental brane \cite{randallsundrum2}. For other foundational papers on extra-dimensional theories, see \cite{antoniadisnewdimattev, newdimatmillimeter, exotickkmodels, gibbonswiltshire, pregeometryakama}.

  Domain walls were first proposed as a way to dynamically localize fields onto a lower dimensional subspace by Rubakov and Shaposhnikov \cite{rubshapdwbranes}. It has been shown that such topological defects can in fact dynamically localize gravity \cite{rsgravitydaviesgeorge2007}, yielding a dynamical realization of the RS2 model. This is attractive from an aesthetic point of view since the brane is no longer a fundamental object. In domain-wall braneworld models, chiral fermions are localized to the wall by Yukawa coupling to the singlet scalar field which contains the domain-wall kink, and scalars are similarly localized through quartic interactions with the kink. Gauge bosons are notoriously difficult to localize, and the only known mechanism which is physically plausible is the Dvali-Shifman mechanism \cite{dsmech}, where the bulk is in the confining phase and respects a gauge group $G$ which is broken by some scalar field on the wall to a subgroup $H$ whose gauge bosons are then repelled by the bulk via confinement dynamics. 
 
  Davies, George and Volkas proposed a phenomenologically plausible domain-wall braneworld model in 4+1D spacetime in which the bulk gauge symmetry is $SU(5)$ which is broken in the usual way to the Standard Model on the wall by an adjoint scalar field \cite{firstpaper}. A consequence of the presence of this $SU(5)$-breaking field in the background along with the domain-wall kink is that the profiles of the SM fermions are in general displaced from the center of the wall and are further split from their $SU(5)$ multiplet partners, leading to a natural realization of the split fermion mechanism \cite{splitfermions}. The authors then showed that in this particular model, the SM fermions could be split appropriately to generate the fermion mass hierarchy naturally \cite{su5branemassfittingpaper}. It was shown in the same paper that generation of the quark mixing angles looked promising, and that proton decay could be suppressed by giving the colored Higgs scalar a large displacement from the brane. However, it was argued at the end of the paper that lepton mixing could not be naturally generated in the model while simultaneously generating the correct charged fermion mass spectra. This motivates us to use additional physics to solve the lepton mixing problem, and in particular to consider the addition of discrete flavor symmetries. 

  Models with discrete flavor symmetries such as $A_{4}$ represent an interesting approach to explaining the quark and lepton mixing patterns. Models of this type were first explored in \cite{maquarka4, emarajasekaran2001a4paper} and references \cite{altarelliferugliosusya4model, volkasa4paper2006} will prove to be relevant for our analysis. In the simplest model using $A_{4}$ in 3+1D with just the Standard Model gauge group \cite{volkasa4paper2006}, typically the different mixing patterns are explained due to $A_{4}$ being spontaneously broken to different subgroups in each sector: $A_{4}\rightarrow{}\mathbb{Z}_{3}$ in the charged fermion sector, and $A_{4}\rightarrow{}\mathbb{Z}_{2}$ in the neutrino sector. This is typically achieved by the addition of two $A_{4}$ triplet Higgs fields which couple to different sectors and which attain different vacuum expectation value (VEV) patterns. When this is done, the CKM matrix is found to be close to the identity and the PMNS matrix assumes a tribimaximal form. When interactions between the two $A_{4}$ triplet flavons are switched off, this arrangement is valid since the two non-aligned VEVs are both global minima of the potentials for each flavon. However, when interactions are switched on the two VEV patterns tend to align and thus the responsible cubic and quartic coupling constants have to be fine-tuned significantly to be small. This problem is known as the vacuum realignment problem, and is typical of theories with discrete flavor symmetries which have extended Higgs sectors of this type. There are in general three ways to ensure that the troublesome interactions are suppressed. 

  One is to make the theory supersymmetric so that the undesired terms are forbidden by holomorphy and renormalization constraints on the superpotential \cite{volkasa4paper2006}. Another is to use additional discrete symmetries forbidding the interactions \cite{grouptheoryvevalignment}. Yet another is to exploit the physics of extra dimensions, by localising the flavons on different branes or by splitting their extra-dimensional profiles with very little overlap so that the interactions are naturally eliminated or very suppressed \cite{feruglioalterellivevalignment, feruglioorbifolda4, kaddoshpallante}. 

  Given that extra-dimensional models have been very successful at explaining the hierarchy problem and can ameliorate one of the major problems of discrete flavor symmetry models, and that discrete flavor symmetries can reproduce realistic leptonic mixing patterns which can be difficult to produce in extra-dimensional models, the combination of the two approaches is quite attractive. There have already been many models in the literature uniting the two approaches, particularly with regards to the warped RS1 scenario. Altarelli and Feruglio first proposed a model based on $A_{4}$ with an SM gauge group and the flavons restricted to different branes \cite{feruglioalterellivevalignment}. There have also been models with GUTs \cite{feruglioorbifolda4, a4symda6dsu5susygut, skinga4su5paper} as well as models with more complicated discrete flavor groups such as the double cover of $A_{4}$, $T'$ \cite{muchenmahanthappayutprimemodel}. It was also shown by Kadosh and Pallante that the flavons could be put into the bulk to allow enough cross-talk between the flavons to generate small quark mixing angles while at the same time maintaining the desired vacuum alignment \cite{kaddoshpallante}. It is also interesting to note that some 3+1D models with flavons in the $1'$ and $1''$ representations of $A_{4}$, which our model also contains and were not previously considered in discrete flavor symmetry models, were proposed in \cite{sking1prime1dprimeflavonpaper, skinga4su5paper}. One of these \cite{skinga4su5paper} was also based on $SU(5)\times{}A_{4}$.

   In this paper, we extend the $SU(5)$ 4+1D domain-wall braneworld model of Davies, George and Volkas \cite{firstpaper} with the inclusion of a discrete $A_{4}$ flavor symmetry group.\footnote{See \cite{takahashisplitseesaw} for another extra-dimensional theory involving flavor symmetry.} We first set up the background configuration with a singlet scalar field forming a domain-wall kink and an adjoint scalar field which attains a lump-like profile which breaks $SU(5)$ to the Standard Model in the middle of the wall in order to facilitate the Dvali-Shifman mechanism. We then dynamically localize the required fermions and flavon Higgs fields embedded in appropriate $SU(5)\times{}A_{4}$ representations to the wall via Yukawa and quartic coupling to the kink-lump solution respectively, and we give the forms of the profiles for the resultant localized SM components which are split according to their hypercharges, yielding a natural realization of the split fermion mechanism.  We show that the results in \cite{su5branemassfittingpaper} with regard to the fermion mass hierarchy problem can be reproduced as well as quark mixing, neutrino mass squared differences and a tribimaximal lepton mixing matrix from a set of 5D Yukawa parameters which are all of the same order of magnitude. In our model, it turns out the required scale of the breaking of $A_{4}$ by the triplet flavons can be altered due to the fact that the Dirac masses for the neutrinos can be suppressed by the split-fermion mechanism, and these scales can vary from the electroweak scale all the way up to the GUT scale. We finally show that splitting the charged $A_{4}$-triplet flavon from the gauge singlet $A_{4}$-triplet flavon can exponentially suppress the interactions responsible for the vacuum realignment problem. 

   In the next section we outline the basic background kink-lump configuration formed from a singlet scalar field which condenses to form the domain wall and an adjoint scalar field which attains a non-zero vacuum expectation value on the wall breaking $SU(5)$ to the Standard Model. Section \ref{sec:mattercontent} outlines both the fermionic and scalar matter content of our model as well as the $SU(5)\times{}A_{4}$ representations to which they are assigned. Section \ref{sec:matterlocalisation} and Sec. \ref{sec:higgslocalisation} then address the dynamical localization of the fermionic matter and the Higgs flavon scalars respectively. Section \ref{sec:ewyukawa} gives details of the electroweak Yukawa Lagrangian of the model and the forms of the fermion mass matrices that arise after the $A_{4}$-triplet flavons condense with the desired vacuum alignment. Our parameter fitting analysis yielding the desired fermion mass spectra, quark mixing, tribimaximal lepton mixing and the correct neutrino mass squared differences is given in Sec.$\:$\ref{sec:examplesolutionmasshierachy}. In Sec. \ref{sec:vacuumrealignment} we discuss our solution to the vacuum realignment problem in our model, with the full flavon interaction potentials given in Appendix \ref{appendix:flavonpotential}. Section \ref{sec:conclusion} is our conclusion.

\section{The Background $SU(5)$ Domain-Wall Braneworld Configuration}
\label{sec:backgrounddw}

 In this section, we briefly cover the basic set-up of the model discussed in \cite{firstpaper, su5branemassfittingpaper}. In our RS2-like model, the extra-dimension $y$ is infinite and the brane to which matter is trapped is formed as a domain wall. To form a domain wall, we need to introduce a singlet scalar field with a $\mathbb{Z}_{2}$-symmetric quartic potential with two discrete and disconnected vacua, and thus find a solution in which the scalar field interpolates between these vacua from $y=-\infty{}$ to $y=+\infty{}$. 

 Gauge bosons are notoriously difficult to localize on domain walls, and they cannot be treated in the same way as fermions and scalars. Instead we conjecture that the Dvali-Shifman mechanism \cite{dsmech} works in 4+1D. In general, the Dvali-Shifman conjecture states that if a gauge group $G$ is confining in the bulk but is spontaneously broken to a subgroup $H$ on the wall, then there should be a mass gap between the glueballs of $G$ and those of $H$ and thus localising $H$-bosons to the wall. For our model, $G=SU(5)$ and $H=SU(3)_{c}\times{}SU(2)_{I}\times{}U(1)_{Y}$, with the appropriate breaking achieved by an adjoint scalar field. 

 Hence, the field content for the background is 
\begin{equation}
\label{eq:backgroundreps}
\eta{}\sim{}1,  \;{} \chi{}\sim{}24,
\end{equation} 
and the most general $Z_{2}$-symmetric potential $V_{\eta{}\chi{}}$ for these fields may be written as
\begin{equation}
\begin{aligned}
\label{eq:backgroundpotential}
V_{\eta{}\chi{}} &= (c\eta^2-\mu^2_{\chi})Tr(\chi^2)+a\eta{}Tr(\chi^3)+\lambda{}_{1}[Tr(\chi^2)]^2 \\
                 &+\lambda{}_{2}Tr(\chi^4)+l(\eta^2-v^2)^2.
\end{aligned}
\end{equation}

\begin{figure}[h]
\includegraphics[scale=1.0]{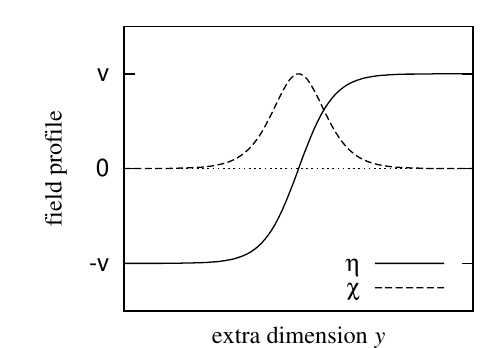}
\caption{A plot of the profiles for $\eta$ and $\chi_{1}$ arising in the background kink-lump solution.}
\label{fig:backgrounddwplot}
\end{figure}

 Finding the requisite background domain wall solution with the desired breaking of $SU(5)$ on the wall involves finding a classical solution dependant solely on the extra-dimensional coordinate $y$ of the Euler-Lagrange equations. To find such a solution, we impose the boundary conditions
 \begin{equation}
\label{eq:backgroundbc}
\eta{}(y=\pm{}\infty{})=\pm{}v, \; \chi{}_{1}(y=\pm{}\infty{})=0,
\end{equation}
where $\chi_{1}$ is the component of $\chi$ corresponding to the appropriately normalised generator, $diag(2/3, 2/3, 2/3, -1, -1)\sqrt{3}/2\sqrt{5}$, proportional to hypercharge, and set all other components of $\chi$ to zero. Numerical solutions exist for a finite region of parameter space. For the sake of simplicity and of yielding an analytic solution, we choose to impose the parameter conditions 
\begin{equation}
\label{eq:analyticdwconditions}
2\mu^{2}_{\chi}(c-\tilde{\lambda})+(2c\tilde{\lambda}-4l\tilde{\lambda}-c^2)v^2=0, \; a=0,
\end{equation}
with $\tilde{\lambda}=\lambda_{1}+\frac{7\lambda_{2}}{30}$. One can then show that the solution is 
\begin{equation}
\label{eq:dwsolution}
\eta{}(y) = v\tanh{(ky)}, \; \chi_{1}(y) = A\sech{(ky)},
\end{equation}
where $k^2=cv^2-\mu{}^{2}_{\chi}$, $A^2=(2\mu^{2}_{\chi}-cv^2)/\tilde{\lambda}$. The solution above is plotted in Fig. \ref{fig:backgrounddwplot}. Numerical solutions still exist for choices outside this parameter region, with the solution for $\eta$ always being kink-like and that for $\chi_{1}$ being lump-like. The stability of configurations such as Eq. \ref{eq:dwsolution} against, for example, formation of non-zero values for other components of $\chi$ has been checked \cite{firstpaper}. 

 One could also consider non-perturbative corrections to the background solution. Outside the wall, the physics of confining $SU(5)$ is fundamentally non-perturbative and cannot easily be calculated. However, since the localization of matter occurs within a distance scale of order $1/k$ we can ignore non-perturbative effects \cite{firstpaper}.

 Before we proceed, it is worth mentioning that there are several scales in our theory. Firstly, since gauge and Yukawa interactions are non-renormalizable in 4+1D, we must impose a cut-off energy scale $\Lambda_{UV}$. In addition to the cut-off scale, we also have $SU(5)$-breaking scale $\Lambda_{SU(5)}=\chi_{1}(y=0)$, the bulk $SU(5)$ confinement scale $\Lambda_{c}$ and finally the inverse width of the domain wall $\Lambda_{DW}=k$. As explained in further detail in \cite{firstpaper}, the required hierarchy for the model to be viable is $\Lambda_{UV}>\Lambda_{SU(5)}>\Lambda_{conf}>\Lambda_{DW}$.

\section{The Matter Content and $A_{4}$ Representations}
\label{sec:mattercontent}

 We now need to introduce three generations of quarks and leptons as well as Higgs fields embedded in representations of $SU(5)\times{}A_{4}$. As is usual for $SU(5)$ grand unified theories (GUTs), the lepton doublets and right-chiral down-type quarks are embedded into $SU(5)$ quintets, while the quark doublets, right-chiral up-type quarks and right-chiral charged leptons are embedded into $SU(5)$ decuplets. Right-chiral neutrinos are introduced as gauge singlets. We will not discuss the group theoretic properties of $A_{4}$ in this paper, see \cite{emarajasekaran2001a4paper, maquarka4, altarelliferugliosusya4model, volkasa4paper2006} for example. 

 In addition to the representations under the gauge group and the discrete flavor symmetry, we must also consider the transformation properties of the fields under the discrete $\mathbb{Z}_2$ reflection symmetry which ensures topological stability of the domain wall. Since interactions which localize fermions to the domain wall are Yukawa interactions of the form $\eta{}\overline{\Psi}\Psi$ and $\eta$ has negative parity, we must have $\overline{\Psi}\Psi \rightarrow{} -\overline{\Psi}\Psi$, which we can satisfy by choosing $\Psi \rightarrow{} i\Gamma{}^{5}\Psi$ or $\Psi \rightarrow{} -i\Gamma^{5}\Psi$. Scalars can have either positive or negative parity. 

 The representations of the fermions, denoted as ($R_1$, $R_2$), where $R_1$ denotes $SU(5)$ representation and $R_2$ denotes the representation under $A_4$, are chosen to be
\begin{equation}
\begin{gathered}
\label{eq:fermionreps}
\Psi_{5} \sim{} (5^*, \,{} 1), \;{} \Psi'_{5} \sim{} (5^*, \,{} 1'), \;{} \Psi''_{5} \sim{} (5^*, \,{} 1'')  \\
\Psi^{i}_{10} \sim{} (10, 1) \; \mathrm{for} \; i=1,2,3   \\
N \sim{} (1, 3)
\end{gathered}
\end{equation}
where $N$ is an $A_{4}$-triplet containing all three right-chiral neutrinos. Under the reflection symmetry, $N \rightarrow{} -i\Gamma{}^{5}N$ and all other fermions transform as $\Psi \rightarrow{} i\Gamma^{5}\Psi$. 

 For the Higgs sector, we require at least one Higgs quintet which contains an electroweak Higgs and a colored Higgs scalar, and some additional flavons as per usual in models with discrete flavor symmetries. Since the three fermion quintets $\Psi_{5}$, $\Psi'_{5}$, and $\Psi''_{5}$ are in the $1$, $1'$, and $1''$ respectively, and since all the fermion decuplets are singlets under $A_4$, to form $A_4$-invariant Yukawa interactions which generate charged lepton and down-type quark masses we similarly require a Higgs quintet under each of the $A_{4}$ representations $1$, $1'$, and $1''$. As all three generations of right-chiral neutrino are embedded into an $A_{4}$ triplet, and since we must form Yukawa interactions involving this triplet and each of the fermion quintets to generate a Dirac neutrino mass matrix, we must have another Higgs quintet in the triplet representation of $A_{4}$. For the desired off-diagonal elements for the Majorana mass matrix for the neutrinos, we also need a gauge singlet Higgs scalar transforming as a triplet under $A_{4}$. Thus, our field content for the Higgs sector can be summarized as 
\begin{equation}
\begin{gathered}
\label{eq:higgsreps}
\Phi \sim{} (5^*, \,{} 1), \;{} \Phi' \sim{} (5^*, \,{} 1'), \;{} \Phi'' \sim{} (5^*, \,{} 1'')  \\
\rho{} \sim{} (5^*, \,{} 3), \;{} \varphi{} \sim{} (1, \,{} 3).
\end{gathered}
\end{equation}

 Under the $\mathbb{Z}_{2}$ reflection symmetry, all scalars except $\varphi$ are chosen to have negative parity, while $\varphi$ is chosen to have positive parity for reasons which will be discussed later in this paper.

\section{Localization of Chiral Fermions}
\label{sec:matterlocalisation}

 To obtain a 3+1D effective field theory on the domain wall and calculate the electroweak Yukawa coupling constants and masses arising in the effective theory, we must localize chiral fermion zero modes embedded in the fermionic fields described in the previous section to the background domain-wall configuration, since these zero modes will be our candidates as the chiral fermions of the SM. This means we must couple the fermion fields to the background fields $\eta$ and $\chi$. 

 Let's consider the right-chiral neutrinos first. Since the right-chiral neutrinos are embedded into $A_{4}$ triplets and since they are gauge singlets, they couple to $\eta$ only and the trapping interaction is simply 
\begin{equation}
\label{eq:neutrinodwyukawa}
Y_{\eta{}\chi{}N} = -h_{1\eta}(\overline{N}N)_{1}\eta{}.
\end{equation}
The 5D Dirac equation that results from this is thus
\begin{equation}
\label{eq:neutrinodiraceq}
i\Gamma{}^{M}\partial_{M}N+h_{1\eta}\eta{}(y)N=0.
\end{equation}

 To examine the effective SM Yukawa interactions for the neutrinos in the effective 4D theory on the wall, we can ignore the Kaluza-Klein (KK) modes and consider only the localized zero mode of the field $N$. We thus can simply look for a solution of the form $N(x,y)=f_{N}(y)\nu_{R}(x)$, where $f_{N}(y)$ is the zero mode profile and $\nu_{R}(x)$ is an $A_{4}$-triplet of 4D massless right-chiral neutrinos satisfying the ansatz 
\begin{equation}
\begin{gathered}
\label{eq:neutrinozeromodeansatz}
i\gamma^{\mu}\partial_{\mu}\nu_{R}(x) = 0, \\
\gamma^{5}\nu_{R}(x) = +\nu_{R}(x).
\end{gathered}
\end{equation} 
Substituting this ansatz into Eq. \ref{eq:neutrinodiraceq}, we find that the profile $f_{N}(y)$ satifies the first order differential equation
\begin{equation}
\label{eq:neutrinoprofilede}
\frac{df_{N}(y)}{dy}+h_{1\eta}v\tanh{(ky)}f_{N}(y) = 0,
\end{equation}
which can be easily solved to yield 
\begin{equation}
\begin{aligned}
\label{eq:neutrinoprofile}
f_{N}(y) &= C_{N}\sech^{\frac{h_{1\eta}v}{k}}{(ky)}, \\ 
         &= \tilde{C}_{N}k^{\frac{1}{2}}\sech^{\tilde{h}_{1\eta}}{(\tilde{y})},
\end{aligned}
\end{equation}
where $\tilde{y}=ky$, $\tilde{h}_{1\eta}=\frac{h_{1\eta}v}{k}$ and $\tilde{C}_{N}=C_{N}k^{-\frac{1}{2}}$ are the non-dimensionalized extra-dimensional coordinate, background Yukawa coupling constant and normalization factor respectively. 

For the analysis in this paper it is convenient to always work with dimensionless variables and functions, thus we define the non-dimensionalized profile $\tilde{f}_{N}(\tilde{y})=k^{-\frac{1}{2}}f_{N}(\tilde{y})$ and normalize it to one to obtain the correct normalization for the effective 4D kinetic term for the zero mode $\nu_{R}$. Since any field increases in mass dimension by half when the dimensionality of spacetime is increased by one, we non-dimensionalize the profiles for any effective 4D mode is the same way.

\begin{figure}[h]
\includegraphics[scale=0.7]{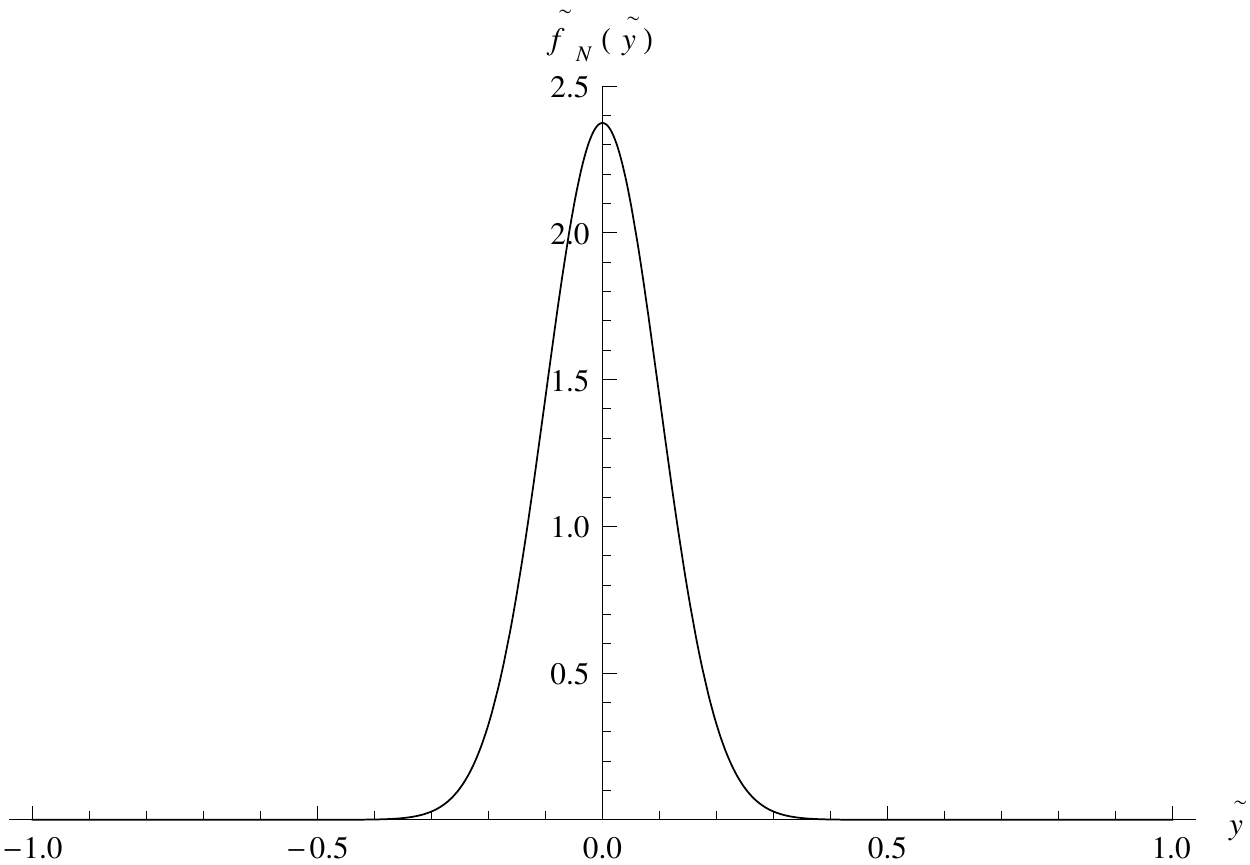}
\caption{The localized right-chiral neutrino triplet profile for the parameter choice $\tilde{h}_{1\eta}=100$.}
\label{fig:neutrinoprofile}
\end{figure}

 As can be seen in Fig. \ref{fig:neutrinoprofile}, the dimensionless profile $\tilde{f}_{N}$ is peaked about $\tilde{y}=0$ and decays exponentially away from the wall. Hence $\nu_{R}$ is indeed localized on the domain wall.

 Next, we consider the fermion quintets. We have one of each of the quintets in the $1$, $1'$, and $1''$ representations of $A_{4}$, which means that due to $A_{4}$-invariance, Yukawa interactions between different generations of the quintets and the background fields $\eta$ and $\chi$ are forbidden. In this case, the coupling of each of these fermions to the background is given by
\begin{equation}
\begin{aligned}
\label{eq:quintetdwyukawa}
Y_{\eta{}\chi{}5} &= h_{5\eta}\overline{\Psi_{5}}\Psi_{5}\eta+h_{5\chi}\overline{\Psi_{5}}\chi^{T}\Psi_{5} \\
                  &+h'_{5\eta}\overline{\Psi'_{5}}\Psi'_{5}\eta+h'_{5\chi}\overline{\Psi'_{5}}\chi^{T}\Psi'_{5} \\
                  &+h''_{5\eta}\overline{\Psi''_{5}}\Psi''_{5}\eta+h''_{5\chi}\overline{\Psi''_{5}}\chi^{T}\Psi''_{5}.
\end{aligned}
\end{equation}
Note the relative minus sign change between the interactions of $\eta$ with $N$ and $\eta$ with the fermion quintets. This choice was made so that a positive $h_{5\eta}$, $h'_{5\eta}$ and $h''_{5\eta}$ correspond to the existence of left-chiral zero modes for the SM components of the respective quintets.

 To find the profiles of these left-chiral zero modes embedded in the quintets, we repeat the analysis done for the field $N$, writing $\Psi^{R}_{5Y}(x,y)=f^{R}_{5Y}(y)\psi^{R}_{5Y}(x)$ for $R=1,\;{}1',\;{}1''$ and $Y=+\frac{2}{3},\;{}-1$, and having the zero modes $\psi^{R}_{5Y}(x)$ satisfy the same ansatz as $\nu_{R}$ given in Eq. \ref{eq:neutrinozeromodeansatz} but with the second condition of right chirality replaced with that of left chirality $\gamma^{5}\psi^{R}_{5Y}(x)=-\psi^{R}_{5Y}(x)$. On substituting the ansatz into the 5D Dirac equation for $\Psi^{R}_{5Y}$,
\begin{equation}
\label{eq:quintetdiraceq}
\Big[i\Gamma^{M}\partial_{M}-h^{R}_{5\eta}\eta{}(y)-\sqrt{\frac{3}{5}}\frac{Y}{2}h^{R}_{5\chi}\chi_{1}(y)\Big]\Psi^{R}_{5Y}(x,y)=0,
\end{equation}
we obtain the ordinary differential equation for the profiles $f^{R}_{5Y}(y)$
\begin{equation}
\label{eq:quintetprofileequation}
\Big[\frac{d}{dy}+h^{R}_{5\eta}v\tanh{(ky)}+h^{R}_{5\chi}A\sqrt{\frac{3}{5}}\frac{Y}{2}\sech{(ky)}\Big]f^{R}_{5Y}(y)=0.
\end{equation}
From the above equation, we find that the non-dimensionalized profiles $\tilde{f}^{R}_{5Y}(\tilde{y})$ of the left-chiral zero modes embedded in the quintets are given by
\begin{equation}
\begin{aligned}
\label{eq:quintetprofiles}
\tilde{f}^{R}_{5Y}(\tilde{y}) &= \tilde{C}^{R}_{5Y}e^{-b^{R}_{5Y}(\tilde{y})}, \\
b^{R}_{5Y}(\tilde{y}) &= \tilde{h}^{R}_{5\eta}\log{\big[\cosh{(\tilde{y})}\big]}+\sqrt{\frac{3}{5}}\tilde{h}^{R}_{5\chi}Y\arctan{\big[\tanh{(\frac{\tilde{y}}{2})}\big]}
\end{aligned}
\end{equation}

\begin{figure}[h]
\includegraphics[scale=0.7]{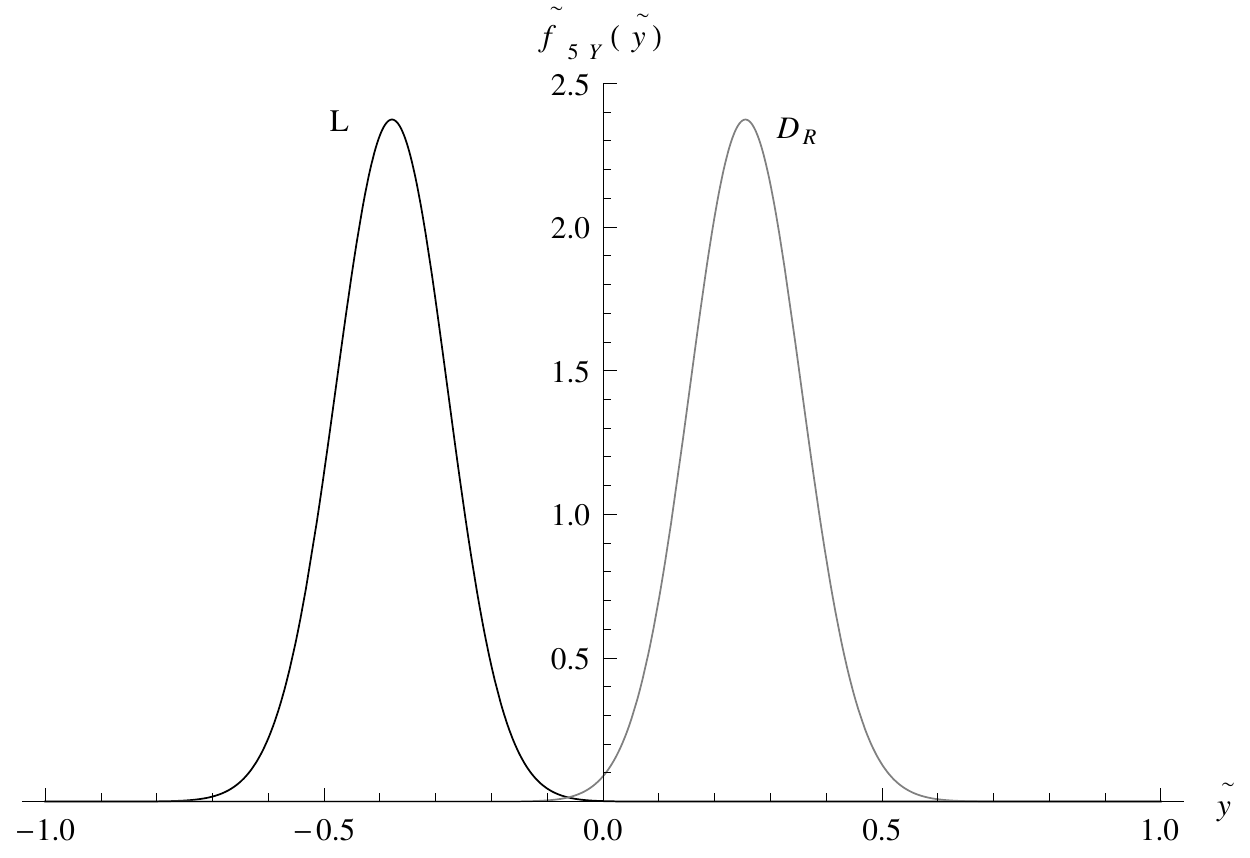}
\caption{The profiles for the localized lepton doublet $L$ and right-chiral down-type quark $D_{R}$ arising from an arbitrary fermion quintet $\Psi_{5}^{R}$ for the parameter choice $\tilde{h}^{R}_{5\eta}=100$ and $\tilde{h}^{R}_{5\chi}=-100$.}
\label{fig:quintetprofiles}
\end{figure}

The consequence of the hypercharge dependency in the coupling to $\chi_{1}$ is that the degeneracy of the profiles for the lepton doublets and the corresponding right-chiral down-type quarks is broken, and furthermore their profiles are displaced from $y=0$, meaning they are \emph{split}. The coupling to $\eta$, $h^{R}_{5\eta}$, roughly determines the widths of the profiles, while the higher the ratio $h^{R}_{5\chi}/h^{R}_{5\eta}$ the more the localization centers of the profiles are displaced from $y=0$. A plot of the profiles for a lepton doublet $L$ and a right-chiral down-type quark $D_{R}$, in any representation $R=1,1',1''$ of $A_{4}$, for the example parameter choice $\tilde{h}^{R}_{5\eta}=100$, $\tilde{h}^{R}_{5\chi}=-100$ is shown in Fig. \ref{fig:quintetprofiles}. 

Finally, we consider the localization of matter embedded in the decuplets $\Psi^{i}_{10}$. Since all of the decuplets are in the trivial representation of $A_{4}$, off-diagonal Yukawa couplings between $\eta$ (or $\chi$) and different generations of $\Psi^{i}_{10}$ are permitted, unlike the case for the fermion quintets. Therefore the most general coupling of the fermion decuplets to the background fields is 
\begin{equation}
\label{eq:decupletdwyukawa}
Y_{\eta{}\chi{}10} = h^{ij}_{10\eta}\overline{\Psi^{i}_{10}}\Psi^{j}_{10}\eta-2h^{ij}_{10\chi}Tr\Big(\overline{\Psi^{i}_{10}}\chi\Psi^{j}_{10}\Big).
\end{equation}

The background Yukawa couplings $h_{10\eta}=\big(h^{ij}_{10\eta}\big)$ and $h_{10\chi}=\big(h^{ij}_{10\chi}\big)$ can be thought of as $3\times{}3$ matrices in the flavor space spanned by the initial 5D fields $\Psi^{i}_{10}$. Due to the Hermiticity of the Lagrangian, both these matrices must be Hermitian and thus we can always choose a basis in which one of them is diagonal at the very least. To be able to pick a basis in which both $h_{10\eta}$ and $h_{10\chi}$ are diagonal requires that they commute, ie. $[h_{10\eta}, h_{10\chi}]=0$. In that case, since $h_{10\eta}=diag(h^{1}_{10\eta}, h^{2}_{10\eta}, h^{3}_{10\eta})$ and $h_{10\chi}=diag(h^{1}_{10\chi}, h^{2}_{10\chi}, h^{3}_{10\eta})$, each generation of decuplet $\Psi^{i}_{10}$ for $i=1,\,{}2,\,{}3$ obeys the Dirac equation
\begin{equation}
\begin{gathered}
\label{eq:decupletdiraceq}
\Big[i\Gamma^{M}\partial_{M}-h^{i}_{10\eta}\eta{}(y)-\sqrt{\frac{3}{5}}\frac{Y}{2}h^{i}_{10\chi}\chi_{1}(y)\Big]\Psi^{i}_{10Y}(x,y)=0, \\
 \mathrm{for} \; Y=-\frac{4}{3}, \,+\frac{1}{3}, \, +2.
\end{gathered}
\end{equation}
Writing $\Psi^{i}_{10Y}(x,y)=f^{i}_{10Y}(y)\psi^{i}_{10Y}(x)$ and again requiring that $\psi^{i}_{10Y}$ is a left-chiral fermion which obeys the massless 4D Dirac equation, we easily find that the non-dimensionalized profiles $\tilde{f}^{i}_{10Y}(y)$ for the decuplet zero modes take the same form as those for the quintets, 
\begin{equation}
\begin{aligned}
\label{eq:decupletprofiles}
\tilde{f}^{i}_{10Y}(\tilde{y}) &= \tilde{C}^{i}_{10Y}e^{-b^{i}_{10Y}(\tilde{y})}, \\
b^{i}_{10Y}(\tilde{y}) &= \tilde{h}^{i}_{10\eta}\log{\big[\cosh{(\tilde{y})}\big]} \\
                       &+\sqrt{\frac{3}{5}}\tilde{h}^{i}_{10\chi}Y\arctan{\big[\tanh{(\frac{\tilde{y}}{2})}\big]}.
\end{aligned}
\end{equation}

\begin{figure}[h]
\includegraphics[scale=0.7]{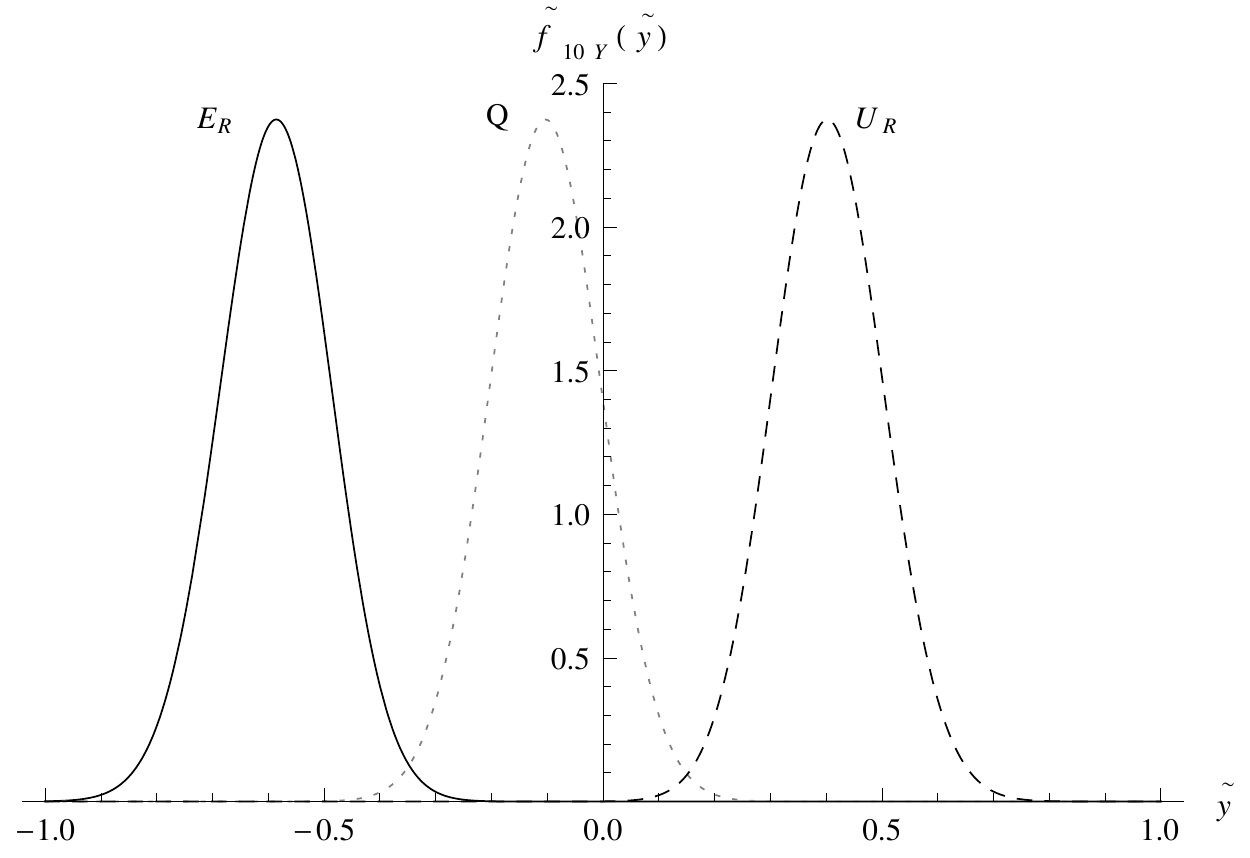}
\caption{Profiles for a right-chiral electron-type lepton $E_{R}$, quark doublet $Q$ and a right-chiral up-type quark $U_{R}$ arising from an arbitrary fermion decuplet $\Psi^{i}_{10}$ for the parameter choice $\tilde{h}^{i}_{10\eta}=100$ and $\tilde{h}^{i}_{10\chi}=100$.}
\label{fig:decupletprofiles}
\end{figure}

 In the case that $h_{10\eta}$ and $h_{10\chi}$ do not commute, then we cannot find a basis in which all three flavors decouple and the flavor diagonal eigenbasis of the full operator $h_{10\eta}+\sqrt{\frac{3}{5}}\frac{Y}{2}h_{10\chi}$ is in fact $y$-dependant. The series of equations becomes a matrix differential equation which is very difficult to solve. This kind of scenario is called the twisted split fermion scenario and has been treated in the context of other models in \cite{twistedsplitfermions, cpandtwistedsf}. We do not consider this case in the analysis and for the sake of simplicity we assume that $h_{10\eta}$ and $h_{10\chi}$ commute in this paper.  A plot of the profiles for the right-chiral electron-type lepton $E_{R}$, quark doublet $Q_{L}$ and right-chiral up-type quark $U_{R}$ for the example parameter choice $\tilde{h}^{i}_{10\eta}=100$, $\tilde{h}^{i}_{10\chi}=100$ for some generation $i$ is shown in Fig. \ref{fig:decupletprofiles}.

 We have successfully shown that fermionic sector of the SM can be localized on the domain-wall brane. Now we must consider the localization of the Higgs scalars and show that electroweak symmetry breaking is possible.

\section{Localization of Higgs fields}
\label{sec:higgslocalisation}

 We now wish to localize the required Higgs scalars on the domain wall. This involves examining the Higgs scalar potential. As is typical of models with discrete flavor symmetries, we have an extended Higgs sector and the full Higgs potential is very complicated. However, most of these interactions are self-interactions amongst the flavons themselves, which do not contribute to the localization of the profiles at leading order. Hence it is sufficient to solely analyze the terms coupling the flavons to $\eta$ and $\chi$ and the bulk masses of the flavons. 

 For the quintet scalars $\Phi=\Phi^{R=1}$, $\Phi'=\Phi^{R=1'}$, $\Phi''=\Phi^{R=1''}$ and the $A_{4}$-triplet $\rho=\Phi^{R=3}$, the localization potentials are easy to write down. They are 
\begin{equation}
\begin{aligned}
\label{eq:quintetscalarlocalisationpotential}
W_{\Phi^{R}} &= \mu^{2}_{\Phi^{R}}(\Phi^{R})^{\dagger}\Phi^{R}+\lambda_{\Phi^{R}\eta}(\Phi^{R})^{\dagger}\Phi^{R}\eta^{2} \\
&+2\lambda_{\Phi^{R}\chi1}(\Phi^{R})^{\dagger}\Phi^{R}Tr\big(\chi^2\big)+\lambda_{\Phi^{R}\chi2}(\Phi^{R})^{\dagger}\big(\chi^{T}\big)^{2}\Phi^{R} \\
&+\lambda_{\Phi^{R}\eta\chi}(\Phi^{R})^{\dagger}\chi^{T}\Phi^{R}\eta, \quad \mathrm{for} \;R=1,\,{}1',\,{}1'', \,{}and\,{}3. 
\end{aligned}
\end{equation}
The mode analysis for the quintets follows that for the Higgs quintet in the original $SU(5)$ braneworld model described in \cite{su5branemassfittingpaper}. Taking the ansatz 
\begin{equation}
\begin{aligned}
\label{eq:quintetscalaransatz}
\Phi^{R}(x,y) &= \sum{}p^{m}_{RY}(y)\phi^{m}_{RY}(x), \\
\Box_{3+1}\phi^{m}_{RY}(x) &= -m_{RY}^{2}\phi^{m}_{RY}(x), 
\end{aligned}
\end{equation}
and substituting it into the resultant 5D Klein-Gordon (KG) equation, one can show that the (non-dimensionalized) profiles for the modes of the Higgs quintets, $\tilde{p}^{m}_{RY}(\tilde{y})$ satisfy a Schr$\ddot{o}$dinger equation with a hyperbolic Scarf potential, $V_{HS}(\tilde{y})$, which can be written as
\begin{equation}
\begin{gathered}
\label{eq:hyperbolicscarfpotential}
-\frac{d^2\tilde{p}^{m}_{RY}}{d\tilde{y}^2}+V_{HS}(\tilde{y})\tilde{p}^{m}_{RY}(y)= E_{RY}\tilde{p}^{m}_{RY}(\tilde{y}), \\
V_{HS}(\tilde{y}) = A^{2}_{RY}+\big(B^{2}_{RY}-A^{2}_{RY}-A_{RY}\big)\sech^2{(\tilde{y})} \\
                  +B_{RY}(2A_{RY}+1)\sech{(\tilde{y})}\tanh{(\tilde{y})},
\end{gathered}
\end{equation}
where $A_{RY}$ and $B_{RY}$ are defined as
\begin{equation}
\begin{aligned}
\label{eq:effectivescalarbackgroundcouplings}
A_{RY} &= \frac{1}{2}\Bigg(-1+\Big(2\big[(\tilde{\lambda}_{\Phi^{R}\chi1}+\frac{3Y^2}{20}\tilde{\lambda}_{\Phi^{R}\chi2}-\tilde{\lambda}_{\Phi^{R}\eta}-\frac{1}{4})^{2} \\
       &+\frac{3Y^2}{20}\tilde{\lambda}^{2}_{\Phi^{R}\eta\chi}\big]^{\frac{1}{2}}-2\tilde{\lambda}_{\Phi^{R}\chi1}-\frac{3Y^2}{10}\tilde{\lambda}_{\Phi^{R}\chi2} \\
       &+2\tilde{\lambda}_{\Phi^{R}\eta}+\frac{1}{2}\Big)^{\frac{1}{2}}\Bigg), \\
B_{RY} &= \frac{\sqrt{\frac{3}{5}}\frac{Y}{2}\tilde{\lambda}_{\Phi^{R}\eta\chi}}{2A_{RY}+1},
\end{aligned}
\end{equation}
the bulk masses, KK mode masses and quartic coupling constants to $\eta$ and $\chi$ are non-dimensionalized as 
\begin{equation}
\begin{gathered}
\label{eq:dimensionlessquinticlambdas}
\tilde{\mu}^{2}_{\Phi^{R}} = \frac{\mu^{2}_{\Phi^{R}}}{k^{2}}, \qquad{} \tilde{m}^{2}_{RY} = \frac{m^{2}_{RY}}{k^{2}}, \\
\tilde{\lambda}_{\Phi^{R}\eta} = \frac{\lambda_{\Phi^{R}\eta}v^{2}}{k^{2}}, \qquad{} \tilde{\lambda}_{\Phi^{R}\chi1} = \frac{\lambda_{\Phi^{R}\chi1}A^{2}}{k^{2}},  \\ \tilde{\lambda}_{\Phi^{R}\chi2} = \frac{\lambda_{\Phi^{R}\chi2}A^{2}}{k^{2}}, \qquad{} \tilde{\lambda}_{\Phi^{R}\eta\chi} = \frac{\lambda_{\Phi^{R}\eta\chi}vA}{k^{2}},
\end{gathered}
\end{equation}
and $E_{RY}$ are the eigenvalues of the above potential, which in terms of the mode masses and fundamental constants in Eq. \ref{eq:dimensionlessquinticlambdas} are
\begin{equation}
\label{eq:hyperbolicscarfeigenvalues}
E_{RY} = \tilde{m}^{2}_{RY}-\tilde{\mu}^{2}_{\Phi^{R}}-\tilde{\lambda}_{\Phi^{R}\eta}+A^{2}_{RY}.
\end{equation}

 The eigenvalues of the hyperbolic Scarf potential are well known \cite{levaishapeinvariantpots, shapeinvkharesukdab, castillohyperscarfpot}. In the case that $A_{RY}>0$, it is known that there exists a set of discrete bound modes for $n=0,1, ... , \lfloor{}A_{RY}\rfloor{}$ with eigenvalues
\begin{equation}
\label{eq:scarfpotentialevalues}
E^{n}_{RY} = 2nA_{RY}-n^2.
\end{equation}

 This gives the mass of the $nth$ localized mode as 
\begin{equation}
\label{eq:quintethiggskkmasses}
\tilde{m}^{2}_{nRY} = \tilde{\mu}^{2}_{\Phi^{R}}+\tilde{\lambda}_{\Phi^{R}\eta}-(A_{RY}-n)^2.
\end{equation}

The lowest energy modes which have the same SM charges as the electroweak Higgs doublet, the $n=0$, $Y=-1$ modes, are the ones we identify as our candidates for the flavons of the effective 4D field theory on the wall. It should be noted that there are regions of parameter space where for a given 5D flavon field, more than one mode has a tachyonic mass. It is also possible to choose parameters such that the modes for the $Y=+2/3$ components, which transform under $SU(3)_c$, would attain tachyonic masses, which would be disastrous since then $SU(3)_c$ would be broken on the wall. Thus, to maintain an unbroken $SU(3)_c$ while employing electroweak symmetry breaking and for the sake of simplicity in the analysis of the electroweak sector, we choose parameters such that only the $n=0$ modes of the electroweak components of $\Phi^{R}$ attain tachyonic masses on the wall while all modes of the colored $Y=+2/3$ components attain positive squared masses.

 It turns out the profiles of the $n=0$ modes of the quintet scalars fields have exactly the same form as for the chiral zero modes for the fermionic quintets described in the previous section, with the $A_{RY}$ playing a role analogous to the $h^{R}_{5\eta}$ and the $B_{RY}$ being analogous to the $h^{R}_{5\chi}$,
\begin{equation}
\begin{aligned}
\label{eq:quintetscalarprofiles}
\tilde{p}_{RY}(\tilde{y}) &= \tilde{C}_{\Phi^{R}Y}e^{-b_{\Phi^{R}Y}(\tilde{y})}, \\
b_{\Phi^{R}Y}(\tilde{y}) &= A_{RY}\log{\big[\cosh{(\tilde{y})}\big]}     \\
                         &+2B_{RY}Y\arctan{\bigg[\tanh{(\frac{\tilde{y}}{2})}\bigg]}.
\end{aligned}
\end{equation}

\begin{figure}[h]
\includegraphics[scale=0.7]{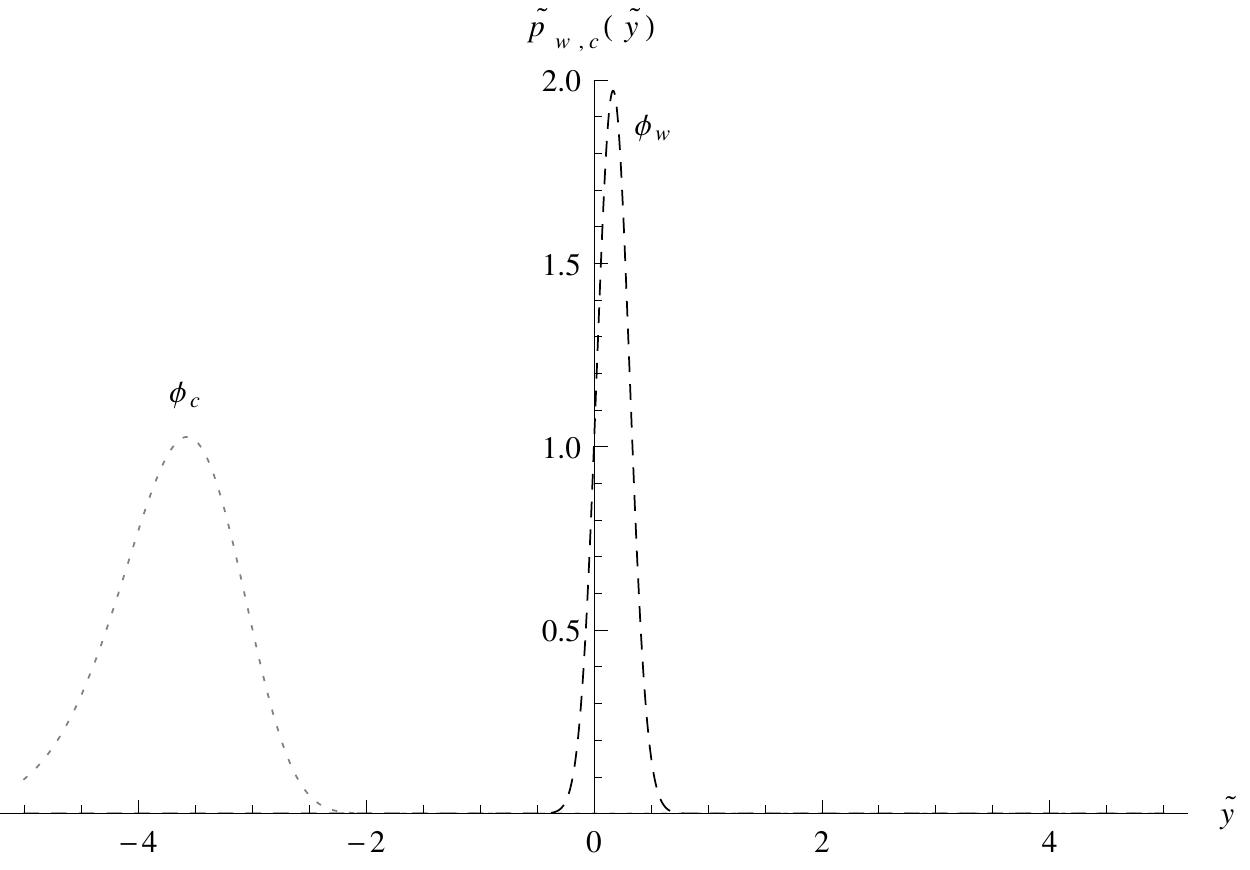}
\caption{The profiles of the localized electroweak Higgs $\phi^{R}_{w}$ and colored Higgs scalar $\phi^{R}_{c}$ for an arbitrary quintet scalar $\Phi^{R}$ for parameters such that $\tilde{\lambda}_{\Phi^{R}\eta}=-7500$, $\tilde{\lambda}_{\Phi^{R}\chi1}= 1500$, $\tilde{\lambda}_{\Phi^{R}\chi2}=-75000$, and $\tilde{\lambda}_{\Phi^{R}\eta\chi}=2000$.}
\label{fig:quintetscalarprofiles}
\end{figure}

From now on, we shall denote the profiles of the $n=0$ modes as $p_{R=1,Y=-1}(y)=p_{w}(y)$, $p_{R=1',Y=-1}(y)=p_{w'}(y)$, $p_{R=1'', Y=-1}(y)=p_{w''}(y)$ and $p_{R=3, Y=-1}(y)=p_{\rho{}w}(y)$ for the electroweak components and the same except with $w$ replaced by $c$ for the $Y=+2/3$ components. Similarly we will denote the corresponding 4D fields for these modes as $\phi_{w,c}(x)$, $\phi_{w',c'}(x)$, $\phi_{w'',c''}(x)$ and $\rho_{w,c}(x)$ for $Y=-1, +2/3$ respectively. A plot of the profiles for the electroweak and colored scalar components of a Higgs quintet in any $A_{4}$ representation is shown in Fig. \ref{fig:quintetscalarprofiles}.

 Now we turn to the gauge singlet, $A_{4}$-triplet scalar $\varphi$. The localization potential for $\varphi$ is given by 
\begin{equation}
\label{eq:phiscalarlocalisationpotential}
W_{\varphi} = \mu^{2}_{\varphi}(\varphi{}\varphi)_{1}+\lambda_{\varphi\eta}(\varphi{}\varphi)_{1}\eta^{2}+2\lambda_{\varphi\chi}(\varphi{}\varphi)_{1}Tr\big(\chi^{2}\big).
\end{equation}
In a similar fashion to the analysis of the quintet scalars, writing down the corresponding Euler-Lagrange equation, then writing $\varphi{}(x,y)=\sum{}p^{m}_{\varphi}(y)\phi^{m}_{\varphi}(x)$ and $\Box_{3+1}\phi^{m}_{\varphi}(x)=-m^2\phi^{m}_{\varphi}(x)$, we find that the modes of $\varphi$ satisfy a Schr$\ddot{o}$dinger equation with a well-known potential, in this case the P$\ddot{o}$schl-Teller potential,
\begin{equation}
\label{eq:phiscalarposchltellerequation}
\Big[-\frac{d^2}{d\tilde{y}^2}+d(d+1)\tanh^{2}{(\tilde{y})}-d\Big]\tilde{p}^{m}_{\varphi}(\tilde{y}) = E^{m}_{\varphi}\tilde{p}^{m}_{\varphi}(\tilde{y}),
\end{equation}
where the parameter $d$ and the eigenvalues $E^{m}_{\varphi}$ are given in terms of the fundamental constants and mode masses as 
\begin{equation}
\begin{aligned}
\label{eq:poschltellerparameters}
d &= \frac{\sqrt{1+4(\tilde{\lambda}_{\varphi{}\eta{}}-\tilde{\lambda}_{\varphi{}\chi})}-1}{2}, \\
E^{m}_{\varphi} &= \tilde{m}^2-\tilde{\mu}^2-\tilde{\lambda}_{\varphi{}\chi}-d.
\end{aligned}
\end{equation}

 Given $d>0$, there exists a tower of discrete localized modes with eigenvalues $E^{n}_{\varphi}=2nd-n^2$, just as before with the hyperbolic Scarf Potentials for the quintet scalars. Similarly, we identify the effective 4D gauge singlet, $A_{4}$-triplet flavon $\varphi_{0}$ as the $n=0$ mode, and we choose parameters such that this mode is the only one which attains a tachyonic mass on the domain-wall brane. The mass squared for this flavon localized to the wall is just
\begin{equation}
\label{eq:varphimasssquared}
\tilde{m}^2_{\varphi_{0}} = \tilde{\mu}^{2}_{\varphi}+\tilde{\lambda}_{\varphi{}\chi}+d,
\end{equation}
 and the profile for $\varphi_{0}$, $\tilde{p}_{\varphi_{0}}(\tilde{y})$ is 
\begin{equation}
\label{eq:varphiprofile}
\tilde{p}_{\varphi_{0}}(\tilde{y}) = \tilde{C}_{\varphi_{0}}\sech^{d}{(\tilde{y})}.
\end{equation}

\begin{figure}[h]
\includegraphics[scale=0.7]{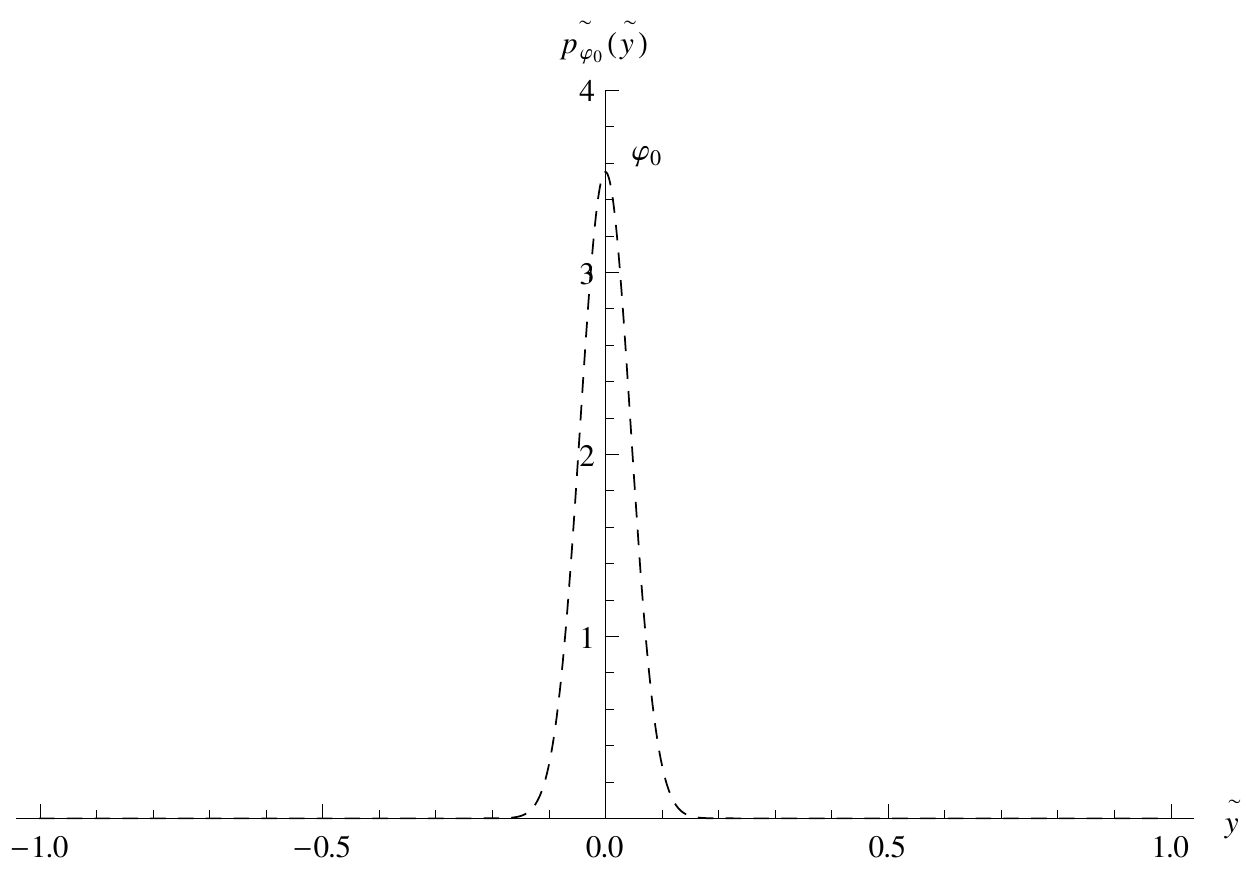}
\caption{The profile of the lowest energy mode, $\varphi_{0}$, of the $A_{4}$ singlet flavon field $\varphi$ for the parameter choice $d=500.00$.}
\label{fig:a4singlethiggsprofile}
\end{figure}

 The lowest energy and tachyonic 4D mode of $\varphi$, $\varphi_{0}$, is always localized at $y=0$. A plot of the profile for the example parameter choice $d=500.00$ is shown in Fig. \ref{fig:a4singlethiggsprofile}. In contrast, the profile of the electroweak component of the other $A_{4}$ triplet, $\rho$, is in general not localized about $y=0$. The natural splitting between the two $A_{4}$ triplets will lead to solutions of the vacuum realignment problem since the splitting will naturally suppress the responsible scalar interactions. This will be covered more extensively in Sec. \ref{sec:vacuumrealignment}.

\section{The Electroweak Yukawa Lagrangian and Fermion Mass Textures}
\label{sec:ewyukawa}

 Given we now have a set of localized modes for the fermions and flavons on the wall, we need to determine the electroweak Yukawa interactions in the model and the effective 4D mass textures after electroweak symmetry breaking. The full 5D electroweak Yukawa potential, $Y_{EW}$, is given by
\begin{equation}
\begin{aligned}
\label{eq:ewyukawalagrangian}
Y_{EW} &= h^{i}_{-}\overline{(\Psi_{5})^{C}}\Psi^{i}_{10}\Phi+h'^{i}_{-}\overline{(\Psi_{5}^{'})^C}\Psi^{i}_{10}\Phi'' \\
       &+h''^{i}_{-}\overline{(\Psi_{5}^{''})^{C}}\Psi^{i}_{10}\Phi'+h^{ij}_{+}\epsilon^{\alpha{}\beta{}\gamma{}\kappa{}\delta}\overline{(\Psi^{i}_{10})_{\alpha{}\beta}^{C}}\Psi^{j}_{10}(\Phi_{\delta})^{*} \\
       &+h_{\rho}\overline{\Psi_{5}}(\rho{}N)_{1}+h'_{\rho}\overline{\Psi'_{5}}(\rho{}N)_{1'}+h''_{\rho}\overline{\Psi''_{5}}(\rho{}N)_{1''} \\
       &+M(\overline{N}N^{C})_{1}+h_{\varphi}\big[(\overline{N}N^{C})_{3s}.\varphi{}\big]_{1}+h.c.
\end{aligned}
\end{equation}

 From this, we can deduce that the elements of the effective 4D up-type quark mass matrix are given by
\begin{equation}
\begin{aligned}
\label{eq:upquarkmassmatrix}
M^{ij}_{U} &= 4h^{ij}_{+}\langle{}\phi_{w}\rangle{}\int{}f_{U^{i}_{R}}(y)f_{Q^{j}}(y)p_{w}(y)\;dy, \\
           &=4\tilde{h}^{ij}_{+}\langle{}\phi_{w}\rangle{}\int\tilde{f}_{U^{i}_{R}}(\tilde{y})\tilde{f}_{Q^{j}}(\tilde{y})\tilde{p}_{w}(\tilde{y})\;d\tilde{y},
\end{aligned}
\end{equation}
where the electroweak Yukawa couplings are non-dimensionalized as $\tilde{h}^{ij}_{+}=h^{ij}_{+}k^{\frac{1}{2}}$. All electroweak Yukawas will be non-dimensionalized this way.
 
 For the down-type quark sector, the rows of the mass matrix are given by
\begin{equation}
\begin{aligned}
\label{eq:downquarkmassmatrix}
M^{1j}_{D} &= \frac{1}{\sqrt{2}}\tilde{h}^{j}_{-}\langle{}\phi_{w}\rangle{}\int{}\tilde{f}_{D_{R}}(\tilde{y})\tilde{f}_{Q^{j}}(\tilde{y})\tilde{p}_{w}(\tilde{y})\;d\tilde{y}, \\
M^{2j}_{D} &= \frac{1}{\sqrt{2}}\tilde{h}'^{j}_{-}\langle{}\phi''_{w}\rangle{}\int{}\tilde{f}_{D'_{R}}(\tilde{y})\tilde{f}_{Q^{j}}(\tilde{y})\tilde{p}_{w''}(\tilde{y})\;d\tilde{y}, \\
M^{3j}_{D} &= \frac{1}{\sqrt{2}}\tilde{h}''^{j}_{-}\langle{}\phi'_{w}\rangle{}\int{}\tilde{f}_{D''_{R}}(\tilde{y})\tilde{f}_{Q^{j}}(\tilde{y})\tilde{p}_{w'}(\tilde{y})\;d\tilde{y}, 
\end{aligned}
\end{equation}
and similarly the columns of the electron-type lepton mass matrix is 
\begin{equation}
\begin{aligned}
\label{eq:electronmassmatrix}
M^{i1}_{E} &= \frac{1}{\sqrt{2}}\tilde{h}^{i}_{-}\langle{}\phi_{w}\rangle{}\int{}\tilde{f}_{E^{i}_{R}}(\tilde{y})\tilde{f}_{L}(\tilde{y})\tilde{p}_{w}(\tilde{y})\;d\tilde{y}, \\
M^{i2}_{E} &= \frac{1}{\sqrt{2}}\tilde{h}'^{i}_{-}\langle{}\phi''_{w}\rangle{}\int{}\tilde{f}_{E^{i}_{R}}(\tilde{y})\tilde{f}_{L'}(\tilde{y})\tilde{p}_{w''}(\tilde{y})\;d\tilde{y}, \\
M^{i3}_{E} &= \frac{1}{\sqrt{2}}\tilde{h}''^{i}_{-}\langle{}\phi'_{w}\rangle{}\int{}\tilde{f}_{E^{i}_{R}}(\tilde{y})\tilde{f}_{L''}(\tilde{y})\tilde{p}_{w'}(\tilde{y})\;d\tilde{y}.
\end{aligned}
\end{equation}

 The Dirac neutrino mass matrix comes from the interaction terms coupling the fermion quintet fields, the neutrino triplet and the Higgs ($5^*$, $3$)-field $\rho$. The form of this mass matrix is then dependent on the vacuum expectation value pattern of the lowest energy mode of $\rho$, $\rho_{w}(x)$. For this paper, we require that $\rho_{w}$ takes the VEV pattern, 
\begin{equation}
\label{eq:rhovevalignement}
\langle{}\rho_{w}\rangle{} = (v_{\rho}, v_{\rho}, v_{\rho}).
\end{equation}
Under this alignment, the form of the Dirac neutrino mass matrix takes the form
\begin{equation}
\begin{aligned}
\label{eq:diracneutrinomassmatrix}
M_{\nu,D} &= \begin{pmatrix} m_{\rho} & m_{\rho} & m_{\rho} \\
                m'_{\rho} & \omega{}m'_{\rho} & \omega^{2}m'_{\rho} \\ 
                m''_{\rho} & \omega^{2}m''_{\rho} & \omega{}m''_{\rho} \end{pmatrix}, \\
              &= \begin{pmatrix} \sqrt{3}m_{\rho} & 0 & 0 \\
                0 & \sqrt{3}m'_{\rho} & 0 \\ 
                0 & 0 & \sqrt{3}m''_{\rho} \end{pmatrix}.U(\omega{}),
\end{aligned}
\end{equation}
where the matrix $U(\omega{})$ is given by
\begin{equation}
\label{eq:uomegamatrix}
U(\omega{}) = \frac{1}{\sqrt{3}}\begin{pmatrix} 1 & 1 & 1 \\
                1 & \omega{} & \omega{}^{2} \\ 
                1 & \omega{}^{2} & \omega{} \end{pmatrix}
\end{equation}
where
\begin{equation}
\begin{aligned}
\label{eq:diracnueigenvalues}
m_{\rho} &= \tilde{h}_{\rho}v_{\rho}\int{}\tilde{f}_{N}(\tilde{y})\tilde{f}_{L}(\tilde{y})\tilde{p}_{\rho{}w}(\tilde{y})\;d\tilde{y}, \\
m'_{\rho} &= \tilde{h}'_{\rho}v_{\rho}\int{}\tilde{f}_{N}(\tilde{y})\tilde{f}_{L'}(\tilde{y})\tilde{p}_{\rho{}w}(\tilde{y})\;d\tilde{y}, \\
m''_{\rho} &= \tilde{h}''_{\rho}v_{\rho}\int{}\tilde{f}_{N}(\tilde{y})\tilde{f}_{L''}(\tilde{y})\tilde{p}_{\rho{}w}(\tilde{y})\;d\tilde{y}.
\end{aligned}
\end{equation}
The matrix $U(\omega{})$ is unitary, and such a factorization of a mass matrix where the mass eigenvalues can be arbitrary while the diagonalization angles are fixed is an example of \emph{form diagonalizability} \cite{lowvolkasformdiagpaper}.

 Lastly, the Majorana neutrino mass matrix is generated from the bare Majorana mass term $M(\overline{N}N^{C})_{1}$ and the Yukawa interaction term $h_{\varphi}\big[(\overline{N}N^{C})_{3s}.\varphi{}\big]_{1}$. The form of the Majorana mass matrix is obviously dependent on the VEV pattern of the lowest energy mode of $\varphi$, $\varphi_{0}$. Unlike for $\rho_{w}$, we instead give the field $\varphi_{0}$ the VEV pattern,
\begin{equation}
\label{eq:varphivevalignment}
\langle{}\varphi_{0}\rangle{} = (0, v_{\varphi}, 0).
\end{equation}
This yields a Majorana neutrino mass matrix of the form,
\begin{equation}
\label{eq:majorananeutrinomassmatrix}
M_{\nu, Maj} = \begin{pmatrix} M & 0 & M_{\varphi} \\
                      0 & M & 0 \\
                      M_{\varphi} & 0 & M \end{pmatrix},
\end{equation}
where 
\begin{equation}
\label{eq:offdiagonalmajoranamassmatrixelementscale}
M_{\varphi} = \tilde{h}_{\varphi}v_{\varphi}\int{}\tilde{f}^{2}_{N}(\tilde{y})\tilde{p}_{\varphi_{0}}(\tilde{y})\; d\tilde{y}.
\end{equation}
Clearly $M^{-1}_{\nu, Maj}$ is diagonalized by the orthogonal matrix 
\begin{equation}
\label{eq:majoranamassmatrixdiagonalizer}
P = \frac{1}{\sqrt{2}}\begin{pmatrix} 1 & 0 & -1 \\
                      0 & \sqrt{2} & 0 \\
                      1 & 0 & 1 \end{pmatrix}.
\end{equation}

 Noting that the effective left Majorana neutrino mass matrix is given by $M_{L}\approx{}-M_{\nu,D}M^{-1}_{\nu, Maj}M^{T}_{\nu,D}$, in the special case that $m_{\rho}=m'_{\rho}=m''_{\rho}$, the left neutrino diagonalization matrix, $V_{\nu{}L}$ assumes the tribimaximal form
\begin{equation}
\begin{aligned}
\label{eq:leftneutrinodiagonalizationmatrix}
V_{\nu{}L} &= U(\omega{})P, \\
           &= \begin{pmatrix} \frac{2}{\sqrt{6}} & \frac{1}{\sqrt{3}} & 0 \\
                              -\frac{\omega{}}{\sqrt{6}} & \frac{\omega{}}{\sqrt{3}} & -\frac{e^{\pi{}i/6}}{\sqrt{2}} \\
                              -\frac{\omega^{2}}{\sqrt{6}} & \frac{\omega^{2}}{\sqrt{3}} & \frac{e^{5\pi{}i/6}}{\sqrt{2}} \end{pmatrix}.
\end{aligned}
\end{equation}

 The PMNS matrix which describes lepton mixing is defined by $V_{PMNS} = V^{\dagger}_{eL}V_{\nu{}L}$, where $V_{eL}$ is the left electron diagonalization matrix. Hence, if we obtain mass textures for the charged leptons such that $V_{eL}\approx{}1$ we recover approximate tribimaximal lepton mixing which is favored by experiment. 

 One also finds that in the special case $m_{\rho}=m'_{\rho}=m''_{\rho}$, the mass eigenvalues for the left Majorana neutrino mass eigenstates are $\frac{-3m^{2}_{\rho}}{M+M_{\varphi}}$, $\frac{-3m^{2}_{\rho}}{M}$, and $\frac{-3m^{2}_{\rho}}{M-M_{\varphi}}$. For the purposes of this paper and for simplicity of analysis, we will choose parameters such that this condition is true. In the next section, we shall show that there exists a non-hierarchical parameter choice such that the Euler angles of the CKM matrix and the charged fermion masses are generated, while at the same time $V_{eL}\approx{}1$, yielding the correct lepton mixing patterns, and that the neutrino mass data can be satisfied in the case that $m_{\rho}=m'_{\rho}=m''_{\rho}$.

\section{Generating the Fermion Mass Hierachy, the CKM Matrix and Lepton Mixing: An Example}
\label{sec:examplesolutionmasshierachy}

 We now give an example parameter choice in which the fermion mass hierarchy, the Euler angles of the CKM matrix, lepton mixing and neutrino mass squared differences are generated. With regard to the CKM matrix, at tree level the CP violating phase has to be put in by hand by giving the electroweak Yukawa coupling constants appropriate phases. We will for the sake of simplicity and clarity of the solution ignore this issue. Note that it is in principle possible for a CP phase in the CKM matrix to be generated from cross-talk between the quark and lepton sectors, since the PMNS matrix given in Eq. \ref{eq:leftneutrinodiagonalizationmatrix} contains non-zero Majorana CP phases. 

 Before we begin the analysis of the fermion mass spectra, let us consider the constraints on the vacuum expectation values of the gauge non-singlet Higgs fields coming from the masses of the $W$ and $Z$ bosons. As is typical with models with discrete flavor symmetries, our model contains an extended Higgs sector. The electroweak Higgs doublets which arise in the effective field theory on the wall are contained in the quintet Higgs flavons, $\Phi$, $\Phi'$, $\Phi''$ and $\rho$. To obtain the required $W$ and $Z$ boson masses, as is usual in a multiple Higgs doublet model, we must have that the sum of the squares of the vacua of these fields is equal to the square of the usual SM Higgs vacuum expectation value of $174\,\mathrm{GeV}$. In our specific model, this requirement is 
\begin{equation}
\label{eq:wzbosonconstraintsonquintetvacua}
\sqrt{\langle{}\phi_{w}\rangle{}^2+\langle{}\phi'_{w}\rangle{}^2+\langle{}\phi''_{w}\rangle{}^2+3v^{2}_{\rho}} = 174\, \mathrm{GeV},
\end{equation}
where $\langle{}\phi_{w}\rangle{}$, $\langle{}\phi'_{w}\rangle{}$ and $\langle{}\phi''_{w}\rangle{}$ are the vacuum expectation values of the electroweak doublets arising from $\Phi$, $\Phi'$ and $\Phi''$ respectively, with $v_{\rho}$ defined as before. For the sake of simplicity in our analysis, we will assume that the vacua obey $\langle{}\phi_{w}\rangle{}=\langle{}\phi'_{w}\rangle{}=\langle{}\phi''_{w}\rangle{}=v_{\rho}=(174\;\mathrm{GeV})/\sqrt{6}=71.0\;\mathrm{GeV}$.

 We first analyze the charged fermion sector. Since the charged fermion mass matrices are derived from Yukawa interactions with the Higgs fields $\Phi$, $\Phi'$, and $\Phi''$, we first need to localize the electroweak components of these Higgs fields. For the purpose of this analysis, we choose parameters such that $A_{R,Y=-1}=100$ and $B_{R,Y=-1}=-10$, for $R=1,\,1',\,1''$. There exists a large region of parameter space spanned by the quartic coupling constants $\tilde{\lambda}_{\Phi^{R}\eta}$, $\tilde{\lambda}_{\Phi^{R}\chi1}$, $\tilde{\lambda}_{\Phi^{R}\chi2}$, $\tilde{\lambda}_{\Phi^{R}\eta\chi}$ and the 5D Higgs squared masses $\mu^{2}_{\Phi^{R}}$ which yields  $A_{R,Y=-1}=100$ and $B_{R,Y=-1}=-10$. Furthermore, a subset of this parameter space is phenomenologically acceptable, namely that at least the lowest energy mode for the electroweak components attains a tachyonic mass on the wall, inducing electroweak symmetry breaking and for which all modes for the colored Higgs components have positive squared masses leaving $SU(3)_{c}$ intact. There also exist choices in this parameter region which satisfy all these constraints and which displace the colored Higgs well away from the wall, suppressing colored-Higgs-induced proton decay. An example from this parameter space is $\tilde{\lambda}_{\Phi^{R}\eta}=-39725$, $\tilde{\lambda}_{\Phi^{R}\chi1}=-24396$, $\tilde{\lambda}_{\Phi^{R}\chi2}=-1.6886\times{}10^{5}$, $\tilde{\lambda}_{\Phi^{R}\eta\chi}=5189.8$, and $\mu^{2}_{\Phi^{R}}=49700$, which gives $\tilde{m}^{2}_{n=0,R,Y=-1}=-25.0$, $\tilde{m}^{2}_{nR,Y=-1}>0$ for all $n>0$, positive squared masses for all colored Higgs modes, and yields $A_{R,Y=2/3}=9.86$ and $B_{R,Y=2/3}=64.6$, which in turn produce a profile localized two characteristic lengths $1/k$ to the left of the center of the domain wall for the lowest energy mode of the colored Higgs. 

 Now that we have localized the electroweak Higgs flavons responsible for charged fermion masses after electroweak symmetry breaking, we can proceed to choose values for the background domain wall Yukawa coupling constants and electroweak Yukawa constants to generate the correct charged fermion mass spectra. We find that substituting the parameter choices\footnote{The appearance of dimensionless numbers of order 100 should not be taken to imply any loss of perturbativity in the Yukawa sector of the low-energy effective theory.  These large numbers are produced by taking the ratio of a given dimensionful quantity with a smaller one, where the denominator has been chosen as a convenient reference point.  To evaluate the validity of perturbation theory, one should look instead at the effective low-energy theory for the quantised field fluctuations about our classical domain wall background solution.  The low-lying modes are the discrete 3+1d KK excitations, and the strength of their Yukawa interactions are given by the effective 3+1d Yukawa coupling constants we have computed.  By construction, the effective 3+1d electroweak Yukawa coupling constants have exactly the same values as those in the standard model, and are therefore perturbative.  Since the underlying 4+1d theory is non-renormalisable, the lack of UV completeness will eventually manifest through the appearance of strong coupling, and the likely loss of unitarity, above some high energy scale.  This breakdown scale has not been computed, but it must be higher than the threshold for the continuum KK modes.  The UV cutoff must be set low enough to avoid the likely pathologies, and some UV completion is, of course, implicitly assumed to exist.},
\begin{equation}
\begin{gathered}
\label{eq:decupletbackgroundyukawachoice}
\tilde{h}^{1}_{10\eta}=701.41, \qquad{} \tilde{h}^{1}_{10\chi}=304.55, \\
\tilde{h}^{2}_{10\eta}=609.43, \qquad{} \tilde{h}^{2}_{10\chi}=263.73, \\
\tilde{h}^{3}_{10\eta}=500.00, \qquad{} \tilde{h}^{3}_{10\chi}=188.63, 
\end{gathered}
\end{equation}
for the fermion decuplet background Yukawa coupling constants
\begin{equation}
\begin{gathered}
\label{eq:fermionquintetbackgroundyukawachoice}
\tilde{h}_{5\eta}=117.48, \qquad{} \tilde{h}_{5\chi}=-239.30, \\
\tilde{h}'_{5\eta}=185.40, \qquad{} \tilde{h}'_{5\chi}=-274.64, \\
\tilde{h}''_{5\eta}=203.50, \qquad{} \tilde{h}''_{5\chi}=-254.82, 
\end{gathered}
\end{equation}
for the quintet background couplings, and
\begin{equation}
\label{eq:decupletelectroweakyukawachoice}
\begin{pmatrix} \tilde{h}^{11}_{+} && \tilde{h}^{12}_{+} && \tilde{h}^{13}_{+} \\
                \tilde{h}^{21}_{+} && \tilde{h}^{22}_{+} && \tilde{h}^{23}_{+} \\
                \tilde{h}^{31}_{+} && \tilde{h}^{32}_{+} && \tilde{h}^{33}_{+} \end{pmatrix}
                = \begin{pmatrix}   640.51  &&  504.28 &&  580.59 \\
                                    501.22  &&  481.66  &&  524.49 \\
                                    129.87  &&  128.95  &&  431.03 \end{pmatrix},
\end{equation}
\begin{equation}
\label{eq:quintetelectroweakyukawachoice}
\begin{pmatrix} \tilde{h}^{1}_{-} && \tilde{h}^{2}_{-} && \tilde{h}^{3}_{-} \\
                \tilde{h}'^{1}_{-} && \tilde{h}'^{2}_{-} && \tilde{h}'^{3}_{-} \\
                \tilde{h}''^{1}_{-} && \tilde{h}''^{2}_{-} && \tilde{h}''^{3}_{-} \end{pmatrix}
                = \begin{pmatrix}   800.00  &&  119.50  &&  119.00 \\
                                    119.50  &&  800.00  &&  119.50 \\
                                    119.00  &&  119.50  &&  800.00 \end{pmatrix},
\end{equation}
for the charged fermion electroweak Yukawa couplings, we obtain the following mass matrices for the up-type quarks, down-type quarks, and electron-type leptons, all in units of MeV, respectively,
\begin{equation}
\label{eq:resultantupquarkmassmatrix}
M_{U} = \begin{pmatrix}  494.21 &&  1044.4   &&  8037.6   \\
                         1044.4 &&  2389.3   &&  14475   \\
                         8037.2 &&  14474   &&   1.7139\times{}10^{5}  \end{pmatrix},                         
\end{equation}
\begin{equation}
\label{eq:resultantdownquarkmassmatrix}
M_{D} = \begin{pmatrix}  4.6631 &&  1.0987   &&  3.2390   \\
                         10.408 &&  116.60   &&  56.779   \\
                         138.17 &&  218.91   &&  4245.2   \end{pmatrix},                         
\end{equation}
\begin{equation}
\label{eq:resultantelectronquarkmassmatrix}
M_{E} = \begin{pmatrix}  0.53478 &&  15.847   && 210.19    \\
                         8.1121\times{}10^{-2} && 107.52    && 217.48    \\
                         5.6198\times{}10^{-2} && 13.762    && 1750.9    \end{pmatrix}.                         
\end{equation}
The mass eigenvalues resulting from these mass matrices are 
\begin{equation}
\label{eq:upcharmtopmasses}
m_{u} = 2.49\;{}\mathrm{MeV}, \quad{}m_{c} = 1.27\;{}\mathrm{GeV}, \quad{}m_{t} = 173\;{}\mathrm{GeV},
\end{equation}
for the up, charm and top quarks,
\begin{equation}
\label{eq:downstrangebottommasses}
m_{d} = 4.47\;{}\mathrm{MeV}, \quad{}m_{s} = 114\;{}\mathrm{MeV}, \quad{}m_{b} = 4.25\;{}\mathrm{GeV},
\end{equation}
for the down, strange and bottom quarks, and
\begin{equation}
\label{eq:electronmuontauonmasses}
m_{e} = 0.511\;{}\mathrm{MeV}, \quad{}m_{\mu{}} = 106\;{}\mathrm{MeV}, \quad{}m_{\tau{}} = 1.78\;{}\mathrm{GeV},
\end{equation}
for the electron, muon and tauon respectively. These results obviously are in agreement within error of current data on charged lepton masses \cite{pdg2010}.

 With regards to quark mixing, the Euler angles, $\Theta^{UL}_{12}$, $\Theta^{UL}_{13}$, and $\Theta^{UL}_{23}$ of the left up quark diagonalization matrix, $V_{UL}$, derived from the up quark mass matrix in Eqn. \ref{eq:resultantupquarkmassmatrix} are approximately
\begin{equation}
\label{eq:euleranglesvuleft}
\Theta^{UL}_{12}=17.264^{\circ{}}, \quad{}\Theta^{UL}_{13}=2.6874^{\circ{}}, \quad{}\Theta^{UL}_{23}=4.8658^{\circ{}},
\end{equation}
and the corresponding angles $\Theta^{DL}_{12}$, $\Theta^{DL}_{13}$, and $\Theta^{DL}_{23}$ of the down quark left diagonalization matrix, $V_{DL}$, are approximately
\begin{equation}
\label{eq:euleranglesvuleft}
\Theta^{DL}_{12}=4.2387^{\circ{}}, \quad{}\Theta^{DL}_{13}=1.8634^{\circ{}}, \quad{}\Theta^{DL}_{23}=2.9746^{\circ{}},
\end{equation}
from which one derives that the Euler angles $\Theta^{CKM}_{12}$, $\Theta^{CKM}_{13}$, and $\Theta^{CKM}_{23}$ of the CKM matrix, given $V_{CKM} = V_{UL}V^{\dagger}_{DL}$, are
\begin{equation}
\label{eq:ckmmixingangles}
\Theta^{CKM}_{12}=13.0^{\circ{}}, \quad{}\Theta^{CKM}_{13}=0.201^{\circ{}}, \quad{}\Theta^{CKM}_{23}=2.39^{\circ{}}.
\end{equation}
These results are in agreement with the current data on quark mixing angles \cite{pdg2010}.

 Finally, the Euler angles $\Theta^{eL}_{12}$, $\Theta^{eL}_{13}$, and $\Theta^{eL}_{23}$ of the left electron diagonalization matrix, $V_{eL}$, are indeed small
\begin{equation}
\begin{gathered}
\label{eq:euleranglesvel}
\Theta^{eL}_{12}=7.32\times{}10^{-2\,{}\circ{}}, \quad{}\Theta^{eL}_{13}=4.15\times{}10^{-3\,{}\circ{}}, \\ 
\Theta^{eL}_{23}=0.925^{\circ{}}.
\end{gathered}
\end{equation}
Thus, the left electron diagonalization matrix is indeed very close to the identity matrix, which means in choosing parameters such that $m_{\rho}=m'_{\rho}=m''_{\rho}$, we get a tribimaximal PMNS matrix generated entirely from the left diagonalization matrix for the neutrinos. The only remaining things to check with regard to the mass fitting are the existence of parameter choices for the right chiral neutrinos generating the desired mass spectra and if, in the case $m_{\rho}=m'_{\rho}=m''_{\rho}$, the experimentally measured neutrino squared mass differences can be satisfied. 

 In the analysis of neutrino masses, we give two example parameter choices for the non-dimensionalized coupling of the right-handed neutrinos to the domain-wall, $\tilde{h}_{1\eta}$, the bare Majorana mass $M$, the non-dimensionalized Yukawa coupling $\tilde{h}_{\varphi}$ describing the strength of the interaction between $N$ and the gauge singlet Higgs field $\varphi$, as well as the quartic coupling parameters determining the localization of the lowest energy modes for the $A_{4}$-triplet flavon fields $\rho$ and $\varphi$. The Dirac masses $m_{\rho}$, $m'_{\rho}$, $m''_{\rho}$ are sensitive to the splittings between the fields $\rho$, the lepton doublets and $N$, and since we still have the freedom to shift $\rho$, we can naturally control the scales of these masses: they can be made to be at the same order as the top quark mass or they can be made to be very light. On the other hand, since the right-handed neutrino triplet $N$ and the flavon $\varphi$ are always localized about $\tilde{y}=0$, the overlap integral $\int{}f^2_{N}(\tilde{y})p_{\varphi_{0}}(\tilde{y})\;d\tilde{y}$ on which the mass scale $M_{\varphi}$ is dependent is always naturally of order 1, so naturally we expect that $M_{\varphi}$ is of the same order as the bare Majorana mass for $N$. 

 In our first examples, we give parameters such that the neutrino Dirac masses are of the order of the electron mass. For the parameter choice 
\begin{equation}
\begin{gathered}
\label{eq:firstneutrinochoice}
\tilde{h}_{1\eta} = 100.00, \\
\tilde{h}_{\rho} = 533.27, \quad{} \tilde{h}'_{\rho} = 267.97, \quad{} \tilde{h}''_{\rho} = 777.57, \\
A_{\rho{}Y=-1}=A_{R=3, Y=-1} = 500.00, \\
B_{\rho{}Y=-1}=B_{R=3, Y=-1} = 380.00,
\end{gathered}
\end{equation}
we get the Dirac masses to be
\begin{equation}
\label{eq:firstdiracnumass}
m_{\rho}=m'_{\rho}=m''_{\rho} = 0.100\; \mathrm{MeV}.
\end{equation}
One can then show that to satisfy the neutrino mass squared differences $\Delta{}m^{2}_{12} = 7.59\times{}10^{-5}\;\mathrm{eV^{2}}$ and $\Delta{}m^{2}_{23} = 2.43\times{}10^{-3}\;\mathrm{eV^{2}}$ from the PDG \cite{pdg2010}, we must have
\begin{equation}
\label{eq:majoranamassscales1}
M = 2.86 \; \mathrm{TeV}, \quad M_{\varphi} = 2.26 \; \mathrm{TeV},
\end{equation}
for which the neutrino masses turn out to be
\begin{equation}
\begin{aligned}
\label{eq:neutrinomasses1}
m_{1} &= \Bigg|\frac{-3m^2_{\rho}}{M+M_{\varphi}}\Bigg| = 5.86\times{}10^{-3} \; \mathrm{eV}, \\
m_{2} &= \Bigg|\frac{-3m^2_{\rho}}{M}\Bigg| = 0.0105 \; \mathrm{eV}, \\
m_{3} &= \Bigg|\frac{-3m^2_{\rho}}{M-M_{\varphi}}\Bigg| = 0.0504 \; \mathrm{eV}, \\
\end{aligned}
\end{equation}
Because the localized mode of $\varphi$, $\varphi_{0}$ is a gauge singlet and $A_{4}$-triplet and thus decoupled from the Standard Model, at least at tree level, we have the freedom to choose the scale of its vacuum expectation value, $v_{\varphi}$. This means we can choose the scale such that the Yukawa coupling $\tilde{h}_{\varphi}$ is of the same order as all the other electroweak Yukawa coupling constants; in this case, if $v_{\varphi}\sim\,10\;{}\mathrm{GeV}$ and $\tilde{h}_{\varphi}\sim{}100$, we get the correct scale for $M_{\varphi}$. 

 Another example we give is one in which the Dirac neutrino masses are the same order as the top quark mass, and the Majorana mass scales are of the order $M_{GUT}=10^{16}\;\mathrm{GeV}$, albeit with $\tilde{h}_{1\eta}$ mildly tuned. For this choice we have,
\begin{equation}
\begin{gathered}
\label{eq:secondneutrinochoice}
\tilde{h}_{1\eta} = 25.000, \\
\tilde{h}_{\rho} = 328.95, \quad{} \tilde{h}'_{\rho} = 208.26, \quad{} \tilde{h}''_{\rho} = 793.17, \\
A_{\rho{}Y=-1}=A_{R=3, Y=-1} = 900.00, \\
B_{\rho{}Y=-1}=B_{R=3, Y=-1} = 600.00,
\end{gathered}
\end{equation}
which gives 
\begin{equation}
\label{eq:seconddiracnumass}
m_{\rho}=m'_{\rho}=m''_{\rho} = 174\; \mathrm{GeV}.
\end{equation}
If we then set 
\begin{equation}
\label{eq:majoranamassscales2}
M = 8.65\times{}10^{15} \; \mathrm{GeV}, \quad M_{\varphi} = 6.85\times{}10^{15} \; \mathrm{GeV},
\end{equation}
we get the masses of the left neutrino mass eigenstates for this parameter choice to be the same as those for the first parameter choice in Eqn. \ref{eq:neutrinomasses1}. Given that the $M_{\varphi}=6.85\times{}10^{15} \; \mathrm{GeV}$, if we have $v_{\varphi}\sim{}10^{13}\,{}\mathrm{GeV}$, then we obtain $\tilde{h}_{\varphi}\sim{}100$ as desired. 

 In summary, we have shown that there exist parameter choices within the model such that the fermion mass hierarchy, light neutrinos and the correct neutrino mass squared differences, quark mixing and a tribimaximal lepton mixing matrix are reproduced from a set of Yukawa coupling constants which are of the same order of magnitude. In light of the recent results of the Daya Bay and RENO neutrino experiments \cite{dayabaytheta13, renotheta13}, which found that $7.9^{\circ{}}<\theta_{13}<9.6^{\circ{}}$ for the PMNS matrix, exact tribimaximal mixing is now excluded. However, a number of assumptions in this analysis were made which ensured an exact tribimaximal form, namely that $m_{\rho}=m'_{\rho}=m''_{\rho}$ and requiring that the left electron diagonalization matrix was close to trivial. It is obvious that deviations from tribimaximal mixing will occur if these assumptions are relaxed, and it is clear that the correct $\theta_{13}$ mixing angle can also be generated in this model. 

 The $m_{\rho}=m'_{\rho}=m''_{\rho}$ assumption generically results in a normal neutrino mass hierarchy, since the only possible hierarchies for this parameter choice are either $m_{1}<m_{2}<m_{3}$, which is the normal hierarchy, or $m_{1}>m_{2}>m_{3}$ which is phenomenologically unacceptable. Breaking this assumption can then lead to inverted or quasidegenerate neutrino mass hierarchies as well as deviations from tribimaximal mixing.

 Like many models based on $A_{4}$, our analysis relied on the alignments of the vacuum expectation values of two $A_{4}$-triplet Higgs fields being different. However, when interactions between the two flavons are switched on, this arrangement is generally destroyed and the VEVs of the two fields align, leading to the vacuum realignment problem. We will discuss the resolution of this problem within our model in the next section.

\section{Resolving the Vacuum Realignment Problem via Splitting}
\label{sec:vacuumrealignment}

 Throughout this paper our analysis has depended on a particular choice of alignments for the vacuum expectation values of the two $A_{4}$-triplet Higgs fields localized to the wall. As per usual, finding valid VEVs for these fields involves finding global minima for the full scalar potential of the theory, which we have put in Appendix \ref{appendix:flavonpotential}. For the gauge charged triplet, $\rho$, we assigned a VEV of the form $(v_{\rho}, v_{\rho}, v_{\rho})$, which induces a breaking of $A_{4}\rightarrow{}\mathbb{Z}_{3}$, and for the neutrally charged triplet flavon $\varphi$ a VEV of the form $(0, v_{\varphi}, 0)$ which breaks $A_{4}$ down to $\mathbb{Z}_{2}$. After these fields are localized and gain tachyonic masses on the wall, one can show that these VEV patterns are indeed valid global minima of the respective self-interaction potentials $V_{\rho}$ and $V_{\varphi}$ given in Eqs. \ref{eq:rhoselfinteraction} and \ref{eq:varphiselfinteraction}. However, when one turns on interactions between the Higgs triplets and the $1$, $1'$ and $1''$ quintet scalars, this is no longer guaranteed \cite{volkasa4paper2006, altarelliferugliosusya4model}. 

 The problem is that once the cross-talking interactions are switched on, minimization of the potential yields a larger number of independent equations than known vacuum expectation values $(v_{\rho}, v_{\varphi}, \langle{}\phi_{w}{}\rangle{}, \langle{}\phi_{w}'{}\rangle{}, \langle{}\phi_{w}''{}\rangle{})$. This means that enforcing the desired vacuum alignment requires an unnatural fine-tuning of the parameters of the scalar potential. The troublesome terms are those which generate soft $A_{4}\rightarrow{}\mathbb{Z}_{2}$ breaking mass terms in the potential for $\rho_{w}$ after $\varphi_{0}$ attains its VEV, and likewise generate soft $A_{4}\rightarrow{}\mathbb{Z}_{3}$ breaking terms for the potential for $\varphi_{0}$ when $\rho_{w}$ condenses. These unwanted interactions involve coupling $1'$, $1''$, $3_{s}$ and $3_{a}$ products of $\rho$ and $\varphi$, for example $(\rho^{\dagger}\rho)_{1'}(\varphi{}\varphi{})_{1''}$ \cite{grouptheoryvevalignment}. In addition to interactions with $\varphi$, there are other interactions generating $A_{4}\rightarrow{}\mathbb{Z}_{2}$ breaking mass terms due to the presence of the $1'$ and $1''$ Higgs quintets $\Phi'$ and $\Phi''$, for example $(\rho^{\dagger}\rho)_{1''}\Phi^{\dagger}\Phi'$. Analogous interactions involving just $\varphi$, $\Phi'$ and $\Phi''$ (and technically $\Phi$ as well) are not problematic since $\varphi$, $\Phi'$ and $\Phi''$ all break $A_{4}$ to the same $\mathbb{Z}_{2}$ subgroups, although there are interactions such as $\Phi''^{\dagger}(\rho\varphi)_{1''}$ which provide corrections to the potential for $\varphi$ as well as that for $\rho$. 

 In our model, we follow the approach in \cite{feruglioalterellivevalignment, feruglioorbifolda4, kaddoshpallante} and split the triplet flavons in the extra dimension to ensure the alignment. Typically, if we suppress the troublesome interactions such that their effective interaction strengths are extremely small, we simply expect a small classical correction to the desired vacua for $\rho_{w}$ of order $\frac{\lambda_{\rho{}\varphi}v^2_{\varphi}}{\lambda{\rho}v_{\rho}}$ for quartic interactions of $\rho$ and $\varphi$ for instance, and similarly for other troublesome mass terms impacting the alignment of $\varphi$. Since $\rho$ is a quintet under $SU(5)$, the electroweak Higgs doublet $\rho_{w}$ arising from it is in general displaced from $\tilde{y}=0$. Since the singlet flavon $\varphi$ is always localized at $\tilde{y}=0$, it is in principle possible to separate the flavons sufficiently so that the operators in $V_{\rho\varphi}$ are naturally suppressed. As previously mentioned, we also have troublesome interactions involving $\rho$ and the $1'$ and $1''$ Higgs quintets. Thus we place the localized mode of $\rho$, $\rho_{w}$, on the opposite side of the wall to those of $\Phi$, $\Phi'$ and $\Phi''$. The interactions coupling all three types of non-trivially $A_{4}$-charged flavons such as $\Phi''^{\dagger}(\rho\varphi)_{1''}$ are naturally suppressed by the splitting of the $A_{4}\rightarrow{}\mathbb{Z}_{3}$ breaking sector generated by $\rho$ from the $A_{4}\rightarrow{}\mathbb{Z}_{2}$ breaking sector generated from $\varphi$, $\Phi'$ and $\Phi''$. 

\begin{figure}[h]
\includegraphics[scale=0.7]{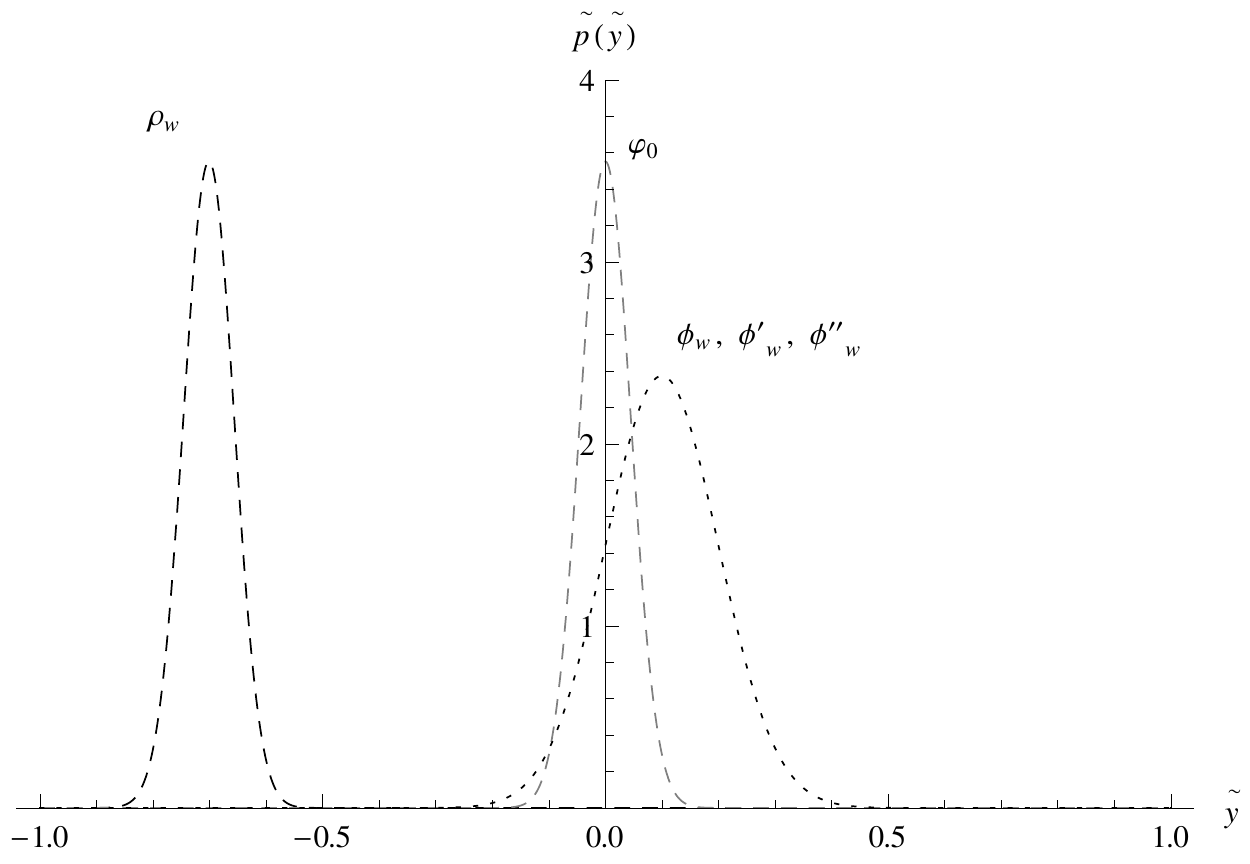}
\caption{Profiles for the localized triplet flavons $\rho_{w}$ and $\varphi_{0}$ as well as those for the $1$, $1'$ and $1''$ Higgs fields $\phi_{w}$, $\phi'_{w}$ and $\phi''_{w}$ for the parameter choices with $A_{\rho{}Y=-1}$ and $B_{\rho{}Y=-1}$ as given in Equation \ref{eq:firstneutrinochoice}, the same localization parameters given for $\Phi$, $\Phi'$ and $\Phi''$ chosen in Sec. \ref{sec:examplesolutionmasshierachy} and $d=500$ for $\varphi$.}
\label{fig:tripletprofilegraph}
\end{figure}

 In our example analysis, for the first choice with $A_{\rho{}Y=-1}$ and $B_{\rho{}Y=-1}$ given in Eq. \ref{eq:firstneutrinochoice}, the same localization parameters for $\Phi$, $\Phi'$ and $\Phi''$ chosen in Sec. \ref{sec:examplesolutionmasshierachy}, we choose $d$, as defined in Eq. \ref{eq:poschltellerparameters} and which determines how localized $\varphi_{0}$ is to the wall, to be set to $d=500.00$. A graph of the profiles for $\rho_{w}$, $\varphi_{0}$ and those of $\phi_{w}$, $\phi'_{w}$, and $\phi''_{w}$ for this parameter choice is shown in Fig. \ref{fig:tripletprofilegraph}. 

 The potentially most troublesome interactions are the cubic interactions of the form $(\rho^{\dagger}\rho)_{3s,a}.\varphi$ and $(\rho^{\dagger}\varphi)_{1, 1', 1''}\Phi^{R=1, 1'', 1'}$, for which the effective 4D couplings are proportional to the integrals $\int{}\tilde{p}^{2}_{\rho{}w}(\tilde{y})\tilde{p}_{\varphi_{0}}(\tilde{y})\;d\tilde{y}$ and $\int{}\tilde{p}_{\rho{}w}(\tilde{y})\tilde{p}_{\varphi_{0}}(\tilde{y})\tilde{p}_{w,w',w''}(\tilde{y})\;d\tilde{y}$ respectively. One has to be careful in calculating the magnitude of the corrections from these interactions since quartic and cubic scalar couplings have different mass dimension; in 4+1D a quartic coupling has mass dimension $-1$ while a cubic coupling has mass dimension $\frac{1}{2}$. For a generic cubic interaction of the types in consideration, if the cubic coupling is $a$, and the quartic self-couplings for a generic $A_{4}$ triplet are of order $\lambda$, provided that the cubic interaction provides just a small perturbation, then we anticipate that this perturbation is proportional to $\frac{a}{\lambda}$. If we choose these numbers to be natural, that is $a=\overline{a}\Lambda^{\frac{1}{2}}_{DW}$ and $\lambda{}=\overline{\lambda{}}\Lambda^{-1}_{DW}$ where $\Lambda_{DW}$ is the UV cut-off for our theory and $\overline{a}$ and $\overline{\lambda{}}$ are dimensionless numbers of order $1$, this means that $\frac{a}{\lambda}\sim{}\Lambda^{\frac{3}{2}}_{DW}$. An overlap integral $O_{3}$ of the profiles of the scalars involved in a cubic interaction is $k^{\frac{1}{2}}$ times the non-dimensionalized integral $\tilde{O}_{3}$, while that for a quartic interaction self-interaction $O_{4}$ is $k\tilde{O}_{4}\sim{}k$. If we now consider $(\rho^{\dagger}\rho)_{3s,a}.\varphi$, after $\rho_{w}$ attains a VEV, this interaction provides a correction linear in $\varphi_{0}$ and so the correction to $v_{\varphi}$ is then of order $\Lambda^{\frac{3}{2}}_{DW}k^{-\frac{1}{2}}\tilde{O}_{3}\frac{v^2_{\rho}}{v^2_{\varphi}}$, and likewise when $\varphi_{0}$ condenses this interaction provides a $A_{4}\rightarrow{}\mathbb{Z}_{2}$ mass term for $\rho_{w}$ which provides a correction $\Lambda^{\frac{3}{2}}_{DW}k^{-\frac{1}{2}}\tilde{O}_{3}\frac{v_{\varphi}}{v_{\rho}}$, where $\tilde{O}_{3}=\int{}\tilde{p}^{2}_{\rho{}w}(\tilde{y})\tilde{p}_{\varphi_{0}}(\tilde{y})\;d\tilde{y}$. We do not know what exactly the scales $k$ and $\Lambda_{DW}$ are, but for the purpose of this analysis we will assume the worst case scenario as far as these corrections are concerned: $k\sim{}1\,\mathrm{TeV}$ and $\Lambda_{DW}\sim{}M_{Planck}=10^{19}\,\mathrm{GeV}$. For this first parameter choice, $\int{}\tilde{p}^{2}_{\rho{}w}(\tilde{y})\tilde{p}_{\varphi_{0}}(\tilde{y})\;d\tilde{y}$ is of the order of $10^{-35}$, which yields the correction to $\langle{}\varphi_{0}\rangle{}$ from this interaction to be of order $10^{-7}v_{\varphi}$, and the correction to $\langle{}\rho_{w}\rangle{}$ to be of order $10^{-10}v_{\rho}$. Likewise, for the interactions of the form $(\rho^{\dagger}\varphi)_{1, 1', 1''}\Phi^{R=1, 1'', 1'}$, the relevant overlap integral $\tilde{O}_{3} = \int{}\tilde{p}_{\rho{}w}(\tilde{y})\tilde{p}_{\varphi_{0}}(\tilde{y})\tilde{p}_{w,w',w''}(\tilde{y})\;d\tilde{y}$ is of order $10^{-31}$ and after the non-triplet gauge charged Higgs fields attain VEVs these interactions give corrections linear in $\varphi_{0}$ and $\rho_{w}$, which turn out to be of the order of $10^{-3}v_{\varphi}$ and $10^{-8}v_{\rho}$ to $\langle{}\varphi_{0}\rangle$ and $\langle{}\rho_{w}\rangle{}$ respectively. 

 With regard to quartic cross-talk interactions, we do not have to worry about scaling of the coupling constants for obvious reasons. For the quartic interactions coupling $\rho$ and $\varphi$, the effective 4D quartic couplings are all proportional to the integral $\int{}\tilde{p}^{2}_{\rho{}w}(\tilde{y})\tilde{p}^2_{\varphi_{0}}(\tilde{y})\;d\tilde{y}$ which for this first parameter choice turns out to be of order $10^{-53}$, yielding corrections of order $(10^{-53}\frac{v^2_{\varphi}}{v^2_{\rho}})v_{\rho}\sim{}10^{-55}v_{\rho}$ to $\langle{}\rho_{w}\rangle{}$ and $(10^{-53}\frac{v^2_{\rho}}{v^2_{\varphi}})v_{\varphi}\sim{}10^{-52}v_{\varphi}$ to $\langle{}\varphi_{0}\rangle{}$. 

 Next, we deal with the quartic interactions coupling $\rho$, $\Phi$ and one of $\Phi'$ or $\Phi''$, which are of the form $(\rho^{\dagger}\rho)_{R=1', 1''}\Phi^{\dagger}\Phi^{R=1'', 1'}$. After dimensional reduction the effective 4D coupling constants for these interactions are proportional to the overlap integral $\int{}\tilde{p}_{\rho{}w}(\tilde{y})\tilde{p}_{w}(\tilde{y})\tilde{p}_{w',w''}(\tilde{y})\;d\tilde{y}$. For this parameter choice these overlap integrals are of order $10^{-22}$, yielding corrections of order $(10^{-22}\frac{\langle{}\phi_{w}\rangle{}\langle{}\phi_{w',w''}\rangle{}v_{\varphi}}{v^3_{\rho}})v_{\rho}\sim{}10^{-22}v_{\rho}$ to $\langle{}\rho_{w}\rangle{}$. 

 Finally, there are quartic interactions coupling both $\rho$ and $\varphi$ to one of $\Phi$, $\Phi'$ or $\Phi''$, which are of the form $\big[(\varphi\varphi)_{3s}.\rho^{\dagger}\big]_{R}\Phi^{R^{*}}$. To leading order, since $(\langle{}\varphi_{0}\rangle{}\langle{}\varphi_{0}\rangle{})_{3s}=((0,v_{\varphi},0).(0,v_{\varphi},0))_{3s}=0$, this interaction does not give a leading order correction to $\langle{}\rho_{w}\rangle{}$, although there are corrections for $\langle{}\varphi_{0}\rangle{}$. After dimensional reduction, the effective 4D coupling strength is proportional to the overlap integral $\int{}\tilde{p}^2_{\varphi_{0}}(\tilde{y})\tilde{p}_{\rho{}w}(\tilde{y})\tilde{p}_{w,w',w''}(\tilde{y})\;d\tilde{y}$, and for the first parameter choice this overlap integral is of the order $10^{-39}$, leading to corrections of order $(10^{-39}\frac{v_{\rho}\langle{}\phi^{R=1,1',1''}_{w}}{v^{2}_{\varphi}})v_{\varphi}\sim{}10^{-37}v_{\varphi}$ to $\langle{}\varphi_{0}\rangle{}$. Obviously any second order corrections to $\langle{}\rho_{w}\rangle{}$ will be much smaller than this. 

 For the second choice with parameters given in Eq. \ref{eq:secondneutrinochoice} with $d=900.00$, we get the value of the relevant overlap integrals to be of order $10^{-50}$ for $\int{}\tilde{p}^{2}_{\rho{}w}(\tilde{y})\tilde{p}_{\varphi_{0}}(\tilde{y})\;d\tilde{y}$, $10^{-42}$ for $\int{}\tilde{p}_{\rho{}w}(\tilde{y})\tilde{p}_{\varphi_{0}}(\tilde{y})\tilde{p}_{w,w',w''}(\tilde{y})\;d\tilde{y}$, $10^{-77}$ for $\int{}\tilde{p}^{2}_{\rho{}w}(\tilde{y})\tilde{p}^2_{\varphi_{0}}(\tilde{y})\;d\tilde{y}$,  $10^{-19}$ for $\int{}\tilde{p}^2_{\rho{}w}(\tilde{y})\tilde{p}^2_{w}(\tilde{y})\tilde{p}_{w',w''}(\tilde{y})\;d\tilde{y}$, and $10^{-54}$ for $\int{}\tilde{p}_{\rho{}w}(\tilde{y})\tilde{p}^{2}_{\varphi_{0}}(\tilde{y})\tilde{p}_{w,w',w''}(\tilde{y})\;d\tilde{y}$. All the corrections to $\langle{}\varphi_{0}\rangle{}$ are not only suppressed by the overlap integrals but by either $\frac{v_{\rho}}{v_{\varphi}}\sim{}10^{-12}$ or its square $(\frac{v_{\rho}}{v_{\varphi}})^2\sim{}10^{-24}$, hence doing a similar analysis to that above shows that the corrections to $\langle{}\varphi_{0}\rangle{}$ are extremely negligible. Doing a similar analysis as above shows that the the overlap integrals are small enough to overcome the ratio $\frac{v_{\varphi}}{v_{\rho}}\sim{}10^{12}$ or its square $(\frac{v_{\varphi}}{v_{\rho}})^2\sim{}10^{24}$ (whichever is relevant) as well as the ratio $\Lambda^{\frac{3}{2}}_{DW}k^{-\frac{1}{2}}$ in the case of the cubic interactions and still ensure that the corrections to $\langle{}\rho_{w}\rangle{}$ are more than several orders of magnitude less than $v_{\rho}$.

\section{Conclusion}
\label{sec:conclusion}

  We found that the $A_4$ extension of the $SU(5)$ 4+1D domain-wall braneworld model of \cite{firstpaper} can generate large mixing angles in the lepton sector. We explicitly demonstrated parameter values that yield tribimaximal lepton mixing with a normal neutrino mass hierarchy together with successful predictions for the hierarchical charged fermion masses and quark mixing angles. Through small departures from this parameter point, the small but nonzero $\theta_{13}$ leptonic mixing angle can be generated, and the neutrino mass spectrum altered to give an inverted hierarchy or a quasi-degenerate pattern. This is a significant extension of the results found in \cite{su5branemassfittingpaper}.

  We also discovered that the troublesome interactions which are responsible for the vacuum realignment problem in analogous 4D models could be suppressed by splitting the profile of $\rho$, which initiates $A_{4}\rightarrow{}\mathbb{Z}_{3}$ breaking, from the profiles of the $A_{4}\rightarrow{}\mathbb{Z}_{2}$ breaking flavons $\varphi$, $\Phi'$ and $\Phi''$. This led to exponential suppression of the overlap integrals of these profiles determining the effective 4D coupling constants for interactions between the $A_{4}$-triplets as well as the troublesome interactions which involve $\Phi'$ and $\Phi''$, ensuring that the contributions of these interactions to the vacua of the localized components of $\rho$ and $\varphi$ were sufficiently small compared to the classical vacua of their respective self-interaction potentials. This maintained the desired vacuum alignments for these flavons required to generate large lepton mixing angles. 

  We have shown in this paper and in \cite{su5branemassfittingpaper} that domain-wall braneworld models with an $SU(5)$ gauge group are generically good for generating the fermion mass hierarchy and quark mixing, and that with the inclusion of the discrete flavor group $A_{4}$ we can attain large lepton mixing angles. Further work along these lines could involve extending either the gauge group (to $SO(10)$ for example \cite{jayneso10paper}, or to even larger groups such as $E_{6}$ \cite{e6domainwallpaper}), the discrete flavor group (for example, to $T'$ which is the double cover of $A_{4}$), or both. Further work could also look at the case where gravity is included or quantum corrections to the fermion mass spectra.

\begin{acknowledgments}

 This work was supported, in part, by the Australian Research Council and the Commonwealth of Australia. We would also like to thank Ferruccio Feruglio for useful discussions as well as the Dipartimento di Fisica 'G. Galilei' at the Universit\`{a} di Padova for its warm hospitality. 

\end{acknowledgments}

\appendix{}
\section{The Higgs Flavon Scalar Interaction Potential}
\label{appendix:flavonpotential}

 In this appendix, we give the potentials describing the interactions amongst different Higgs flavon fields localized to the domain wall. The background scalar field potential $V_{\eta{}\chi{}}$ yielding the background kink-lump solution was given in Eqn. \ref{eq:backgroundpotential} in Sec. \ref{sec:backgrounddw}. The potentials coupling the flavons to the background fields $\eta$ and $\chi$, $W_{\Phi}$, $W_{\Phi'}$, $W_{\Phi''}$, $W_{\rho}$ and $W_{\varphi}$, were given in Sec. \ref{sec:higgslocalisation} in Eqns. \ref{eq:quintetscalarlocalisationpotential} and \ref{eq:phiscalarlocalisationpotential}, which after localization of the scalars and an appropriate choice of parameters contribute tachyonic masses to the effective self-interaction potentials for the lightest localized modes of these fields. Thus let $V_{\rho{}}$, $V_{\varphi{}}$, $V_{\Phi{}}$, $V_{\Phi'{}}$, and $V_{\Phi''{}}$ be potentials containing the quartic self-interactions for each of these fields along with their localization potentials. Let the potentials $V_{\rho\varphi}$, $V_{\rho{}\Phi{}\Phi'{}\Phi''}$, $V_{\varphi{}\Phi{}\Phi'{}\Phi''}$, $V_{\rho\varphi\Phi\Phi'\Phi''}$, and $V_{\Phi\Phi'\Phi''}$  be those containing cross-talk interactions between the fields in the subscripts. Then the full scalar potential of the theory, $V$, is given by
\begin{equation}
\begin{aligned}
\label{eq:fullscalarpotential}
V &= V_{\eta{}\chi{}}+V_{\rho}+V_{\varphi}+V_{\Phi}+V_{\Phi'}+V_{\Phi''}+V_{\rho{}\varphi{}} \\
  &+V_{\rho{}\Phi{}\Phi'{}\Phi''}+V_{\varphi{}\Phi{}\Phi'{}\Phi''}+V_{\rho\varphi\Phi\Phi'\Phi''}+V_{\Phi\Phi'\Phi''},
\end{aligned}
\end{equation}
where
\begin{equation}
\begin{aligned}
\label{eq:rhoselfinteraction}
V_{\rho{}} &= \lambda_{\rho}^{1}(\rho^{\dagger}\rho)_{1}(\rho^{\dagger}\rho)_{1}+\lambda_{\rho}^{2}(\rho^{\dagger}\rho)_{1'}(\rho^{\dagger}\rho)_{1''} \\
           &+\lambda_{\rho}^{3}\big[(\rho^{\dagger}\rho)_{3s}.(\rho^{\dagger}\rho)_{3s}\big]_{1}+\lambda_{\rho}^{4}\big[(\rho^{\dagger}\rho)_{3a}.(\rho^{\dagger}\rho)_{3a}\big]_{1} \\
           &+i\lambda_{\rho}^{5}\big[(\rho^{\dagger}\rho)_{3s}.(\rho^{\dagger}\rho)_{3a}\big]_{1}+W_{\rho} \\
           &=\lambda_{\rho}^{1}(\rho^{\dagger}\rho)_{1}(\rho^{\dagger}\rho)_{1}+\lambda_{\rho}^{2}(\rho^{\dagger}\rho)_{1'}(\rho^{\dagger}\rho)_{1''} \\
           &+\lambda_{\rho}^{3}\big[(\rho^{\dagger}\rho)_{3s}.(\rho^{\dagger}\rho)_{3s}\big]_{1}+\lambda_{\rho}^{4}\big[(\rho^{\dagger}\rho)_{3a}.(\rho^{\dagger}\rho)_{3a}\big]_{1} \\
           &+i\lambda_{\rho}^{5}\big[(\rho^{\dagger}\rho)_{3s}.(\rho^{\dagger}\rho)_{3a}\big]_{1}+\mu^{2}_{\rho}(\rho^{\dagger}\rho)_{1}+\lambda_{\rho{}\eta}(\rho^{\dagger}\rho)_{1}\eta^{2} \\
&+2\lambda_{\rho{}\chi1}(\rho^{\dagger}\rho)_{1}Tr\big(\chi^2\big)+\lambda_{\rho{}\chi2}(\rho^{\dagger}\big(\chi^{T}\big)^{2}\rho{})_{1} \\
&+\lambda_{\rho\eta\chi}(\rho^{\dagger}\chi^{T}\rho)_{1}\eta,
\end{aligned}
\end{equation}

\begin{equation}
\begin{aligned}
\label{eq:varphiselfinteraction}
V_{\varphi} &= \delta_{\varphi}(\varphi{}\varphi{}\varphi{})_{1}+\lambda_{\varphi}^{1}(\varphi{}\varphi{})_{1}(\varphi{}\varphi{})_{1} \\
            &+\lambda_{\varphi}^{2}(\varphi{}\varphi{})_{1'}(\varphi{}\varphi{})_{1''}+\lambda_{\varphi}^{3}\big[(\varphi{}\varphi{})_{3s}(\varphi{}\varphi{})_{3s}\big]_{1}+W_{\varphi} \\
            &= \delta_{\varphi}(\varphi{}\varphi{}\varphi{})_{1}+\lambda_{\varphi}^{1}(\varphi{}\varphi{})_{1}(\varphi{}\varphi{})_{1} \\
            &+\lambda_{\varphi}^{2}(\varphi{}\varphi{})_{1'}(\varphi{}\varphi{})_{1''}+\lambda_{\varphi}^{3}\big[(\varphi{}\varphi{})_{3s}(\varphi{}\varphi{})_{3s}\big]_{1} \\
            &+\mu^{2}_{\varphi}(\varphi{}\varphi)_{1}+\lambda_{\varphi\eta}(\varphi{}\varphi)_{1}\eta^{2}+2\lambda_{\varphi\chi}(\varphi{}\varphi)_{1}Tr\big(\chi^{2}\big),
\end{aligned}
\end{equation}

\begin{equation}
\begin{aligned}
\label{eq:phiphiprimephidoubleprimeselfinteraction}
V_{\Phi^{R}} &= \lambda_{\Phi^{R}}((\Phi^{R})^{\dagger}\Phi^{R})^2+W_{\Phi^{R}} \\
             &= \lambda_{\Phi^{R}}((\Phi^{R})^{\dagger}\Phi^{R})^2+\mu^{2}_{\Phi^{R}}(\Phi^{R})^{\dagger}\Phi^{R}+\lambda_{\Phi^{R}\eta}(\Phi^{R})^{\dagger}\Phi^{R}\eta^{2} \\
&+2\lambda_{\Phi^{R}\chi1}(\Phi^{R})^{\dagger}\Phi^{R}Tr\big(\chi^2\big)+\lambda_{\Phi^{R}\chi2}(\Phi^{R})^{\dagger}\big(\chi^{T}\big)^{2}\Phi^{R} \\
&+\lambda_{\Phi^{R}\eta\chi}(\Phi^{R})^{\dagger}\chi^{T}\Phi^{R}\eta, \quad \mathrm{for} \;R=1,\,{}1',\,{}1'',
\end{aligned}
\end{equation}

\begin{equation}
\begin{aligned}
\label{eq:rhovarphiinteraction}
V_{\rho\varphi} &= \delta^{s}_{\rho\varphi}\big[(\rho^{\dagger}\rho)_{3s}.\varphi\big]_{1}+i\delta^{a}_{\rho\varphi}\big[(\rho^{\dagger}\rho)_{3a}.\varphi\big]_{1} \\
                &+\lambda^{1}_{\rho\varphi}(\rho^{\dagger}\rho)_{1}(\varphi{}\varphi)_{1}+\lambda^{2}_{\rho\varphi}(\rho^{\dagger}\rho)_{1'}(\varphi{}\varphi)_{1''} \\
                &+\lambda^{2*}_{\rho\varphi}(\rho^{\dagger}\rho)_{1''}(\varphi{}\varphi)_{1'}+\lambda^{3}_{\rho\varphi}\Big[(\rho^{\dagger}\rho)_{3s}.(\varphi{}\varphi)_{3s}\Big]_{1} \\
                &+i\lambda^{4}_{\rho\varphi}\Big[(\rho^{\dagger}\rho)_{3a}.(\varphi{}\varphi)_{3s}\Big]_{1}
\end{aligned}
\end{equation}

\begin{equation}
\begin{aligned}
\label{eq:rhophiphiprimephidoubleprime}
V_{\rho{}\Phi{}\Phi'{}\Phi''} &= \lambda^{1}_{\rho{}\Phi}(\rho^{\dagger}\rho)_{1}(\Phi^{\dagger}\Phi)+\lambda_{\rho{}\Phi'}(\rho^{\dagger}\rho)_{1}(\Phi'^{\dagger}\Phi') \\
                              &+\lambda_{\rho\Phi''}(\rho^{\dagger}\rho)_{1}(\Phi'^{\dagger}\Phi')+\lambda_{\rho\Phi\Phi'}(\rho^{\dagger}\rho)_{1''}(\Phi^{\dagger}\Phi') \\
                  &+\lambda^{*}_{\rho\Phi\Phi'}(\rho^{\dagger}\rho)_{1'}(\Phi'^{\dagger}\Phi)+\lambda_{\rho\Phi\Phi''}(\rho^{\dagger}\rho)_{1'}(\Phi^{\dagger}\Phi'')    \\
     &+\lambda^{*}_{\rho\Phi\Phi''}(\rho^{\dagger}\rho)_{1''}(\Phi''^{\dagger}\Phi)+\lambda^{1}_{\rho\Phi'\Phi''}(\rho^{\dagger}\rho)_{1''}(\Phi'^{\dagger}\Phi'')   \\
     &+\lambda^{1*}_{\rho\Phi'\Phi''}(\rho^{\dagger}\rho)_{1'}(\Phi''^{\dagger}\Phi')+\lambda^{2}_{\rho{}\Phi}\big[(\rho^{\dagger}\Phi).(\rho^{\dagger}\Phi)\big]_{1} \\                                                                                                       
                              &+\lambda^{2*}_{\rho\Phi}\big[(\Phi^{\dagger}\rho).(\Phi^{\dagger}\rho)\big]_{1}+\lambda^{2}_{\rho\Phi'\Phi''}\big[(\rho^{\dagger}\Phi').(\rho^{\dagger}\Phi'')\big]_{1} \\
                              &+\lambda^{2*}_{\rho\Phi'\Phi''}\big[(\Phi'^{\dagger}\rho).(\Phi''^{\dagger}\rho)\big]_{1}
\end{aligned}
\end{equation}

\begin{equation}
\begin{aligned}
\label{eq:varphiphiphiprimephidoubleprime}
V_{\varphi{}\Phi{}\Phi'{}\Phi''} &= \lambda_{\varphi\Phi}(\Phi^{\dagger}\Phi)(\varphi{}\varphi)_{1}+\lambda_{\varphi\Phi'}(\Phi'^{\dagger}\Phi')(\varphi{}\varphi)_{1} \\
                           &+\lambda_{\varphi\Phi''}(\Phi''^{\dagger}\Phi'')(\varphi{}\varphi)_{1}+\lambda_{\varphi\Phi\Phi'}(\Phi^{\dagger}\Phi')(\varphi\varphi)_{1''} \\
                   &+\lambda^{*}_{\varphi\Phi\Phi'}(\Phi'^{\dagger}\Phi)(\varphi\varphi)_{1'}+\lambda_{\varphi\Phi\Phi''}(\Phi^{\dagger}\Phi'')(\varphi\varphi)_{1'}    \\
                   &+\lambda^{*}_{\varphi\Phi\Phi''}(\Phi''^{\dagger}\Phi)(\varphi\varphi)_{1''}+\lambda_{\varphi\Phi'\Phi''}(\Phi'^{\dagger}\Phi'')(\varphi\varphi)_{1''} \\
                   &+\lambda^{*}_{\varphi\Phi'\Phi''}(\Phi''^{\dagger}\Phi')(\varphi\varphi)_{1'},
\end{aligned}
\end{equation}

\begin{equation}
\begin{aligned}
\label{eq:rhovarphiphiphiprimephidoubleprimeinteractions}
V_{\rho\varphi\Phi\Phi'\Phi''} &= \delta_{\rho\varphi\Phi}(\varphi\rho^{\dagger})_{1}\Phi+\delta^{*}_{\rho\varphi\Phi}\Phi^{\dagger}(\rho{}\varphi)_{1} \\
                               &+\delta_{\rho\varphi\Phi'}(\varphi\rho^{\dagger})_{1''}\Phi'+\delta^{*}_{\rho\varphi\Phi'}\Phi'^{\dagger}(\rho\varphi)_{1'} \\
                               &+\delta_{\rho\varphi\Phi''}(\varphi\rho^{\dagger})_{1'}\Phi''+\delta^{*}_{\rho\varphi\Phi''}\Phi''^{\dagger}(\rho\varphi)_{1''} \\
   &+\lambda_{\rho\varphi\Phi}\big[(\varphi\varphi)_{3s}.\rho^{\dagger}\big]_{1}\Phi+\lambda^{*}_{\rho\varphi\Phi}\Phi^{\dagger}\big[\rho{}.(\varphi\varphi)_{3s}\big]_{1} \\
   &+\lambda_{\rho\varphi\Phi'}\big[(\varphi\varphi)_{3s}.\rho^{\dagger}\big]_{1''}\Phi'+\lambda^{*}_{\rho\varphi\Phi'}\Phi'^{\dagger}\big[\rho{}.(\varphi\varphi)_{3s}\big]_{1'} \\
   &+\lambda_{\rho\varphi\Phi''}\big[(\varphi\varphi)_{3s}.\rho^{\dagger}\big]_{1'}\Phi''+\lambda^{*}_{\rho\varphi\Phi''}\Phi''^{\dagger}\big[\rho{}.(\varphi\varphi)_{3s}\big]_{1''},
\end{aligned}
\end{equation}

\begin{equation}
\begin{aligned}
\label{eq:phiphiprimephidoubleprimemixedinteraction}
V_{\Phi\Phi'\Phi''} &= \lambda^{1}_{\Phi{}\Phi'{}}(\Phi^{\dagger}\Phi)(\Phi'^{\dagger}\Phi')+\lambda^{2}_{\Phi{}\Phi'{}}(\Phi^{\dagger}\Phi')(\Phi'^{\dagger}\Phi) \\
                    &+\lambda^{1}_{\Phi{}\Phi''{}}(\Phi^{\dagger}\Phi)(\Phi''^{\dagger}\Phi'')+\lambda^{2}_{\Phi{}\Phi''{}}(\Phi^{\dagger}\Phi'')(\Phi''^{\dagger}\Phi) \\
                    &+\lambda^{1}_{\Phi'{}\Phi''{}}(\Phi'^{\dagger}\Phi')(\Phi''^{\dagger}\Phi'')+\lambda^{2}_{\Phi'{}\Phi''{}}(\Phi'^{\dagger}\Phi'')(\Phi''^{\dagger}\Phi') \\
                    &+\lambda^{1}_{\Phi{}\Phi'{}\Phi''{}}(\Phi^{\dagger}\Phi')(\Phi^{\dagger}\Phi'')+\lambda_{\Phi{}\Phi'{}\Phi''{}}^{1*}(\Phi'^{\dagger}\Phi)(\Phi''^{\dagger}\Phi) \\ 
                    &+\lambda^{2}_{\Phi{}\Phi'{}\Phi''{}}(\Phi'^{\dagger}\Phi)(\Phi'^{\dagger}\Phi'')+\lambda_{\Phi{}\Phi'{}\Phi''{}}^{2*}(\Phi^{\dagger}\Phi')(\Phi''^{\dagger}\Phi') \\
                    &+\lambda^{3}_{\Phi{}\Phi'{}\Phi''{}}(\Phi''^{\dagger}\Phi)(\Phi''^{\dagger}\Phi')+\lambda_{\Phi{}\Phi'{}\Phi''{}}^{3*}(\Phi^{\dagger}\Phi'')(\Phi'^{\dagger}\Phi'')
\end{aligned}
\end{equation}

\newpage

\bibliographystyle{ieeetr}
\bibliography{bibliography2.bib}

\end{document}